\documentclass[USenglish, 12pt, fleqn, oneside]{article}
\usepackage[margin=1.15in]{geometry}
\usepackage{setspace}
\usepackage[english]{babel}
\usepackage{times}
\usepackage{abstract}
\usepackage{graphicx}
\usepackage{lscape}
\usepackage{float}
\usepackage{subcaption}
\usepackage{color}
\usepackage{epsfig}
\usepackage{booktabs}
\usepackage{multirow}
\usepackage{appendix}
\usepackage{epsfig}
\usepackage{amsmath,amssymb,amsthm,bm}
\usepackage{url}
\usepackage{accents}
\usepackage{enumerate}
\usepackage{paralist}
\usepackage{natbib}
\usepackage{newlfont}
\usepackage{pstricks,pstricks-add,pst-plot,pst-node}
\usepackage{dsfont}
\usepackage{colortbl}
\bibpunct{(}{)}{;}{a}{,}{,}
\usepackage{fancyhdr} 

\usepackage{amsfonts,amsmath,amssymb,amsthm}
\usepackage{mathtools}

\newcommand{\argmin}{{\operatorname{argmin}}}

\newcommand{\xe}{{\cal X}_e}
\newcommand{\xl}{{\cal X}_\ell}

\newlength{\dhatheight}

\begin{document}
	
	\thispagestyle{empty}
	
	\author{
		Arthur Charpentier\thanks{%
        This work was supported for AC by the NSERC Discovery Grant; under Grant [RGPIN-2019-04190] and the SCOR Foundation for Science.
        } \\
        {\footnotesize Universit\'e du Qu\'ebec \`a Montr\'eal. 405 Rue Sainte-Catherine Est, Montr\'eal, QC H2L 2C4, Canada} \\ [10pt]
        Qiheng Guo\thanks{%
		Corresponding author. Phone: +1-(765)-285-8645.
		E-mail address: qguo@bsu.edu.} \\
        {\footnotesize Ball State University. 2000 W University Avenue. Muncie, IN 47306. USA } \\ [10pt]
        Michael Ludkovski \\
        {\footnotesize University of California, Santa Barbara. Santa Barbara, CA 93106. USA}
	}

	\title{{\large Functional Analysis of Loss-development Patterns in P\&C Insurance}}
	
    \date{September 2025}
	
	{\let\newpage\relax\maketitle}	
	\maketitle

	\singlespacing

	\begin{abstract}                                                                             	
		
		We analyze loss development in NAIC Schedule P loss triangles using functional data analysis methods. Adopting the functional 
        viewpoint, our dataset comprises 3300+ curves of incremental loss ratios (ILR) of workers' compensation lines over 24 accident years. Relying on functional data depth, we first study similarities and differences in development patterns based on company-specific covariates, as well as identify anomalous ILR curves.
        %
        The exploratory findings motivate the probabilistic forecasting framework developed in the second half of the paper.
        We propose a functional model to complete partially developed ILR curves based on partial least squares regression of PCA scores. Coupling the above with functional bootstrapping allows us to quantify future ILR uncertainty jointly across all future lags. We demonstrate that our method has much better probabilistic scores relative to Chain Ladder and in particular can provide accurate functional predictive intervals. 
		\vspace{3mm}
		
		
		
	\end{abstract}
	\thispagestyle{empty}
	
	
	\newpage 

    
\section{Introduction}

\subsection{Motivation and Literature}
    
	A property and casualty (P\&C) insurer sells insurance policies in multiple businesses. The risks, or loss distributions, differ among business lines and spread across time as claims take multiple years to be fully paid off. When aggregating all claims within a company's given business line, paid losses gradually accumulate each year starting from the year claims are filed. Linking paid loss observations across years results in a \emph{loss development} pattern, indexed by accident years and development lags. Studying loss development helps the insurer predict its future paid losses, or loss reserves, a crucial part in its balance sheet. 
	
	Different business lines have different patterns in loss development. Generally speaking, property lines, such as auto physical damage, have a loss development of 1-3 years, while casualty businesses, such as private passenger auto liability and workers' compensation, can have a loss development of 5-30 years. Within each casualty line, the amount of losses in each development lag is  heterogeneous across the industry and business lines. While loss development patterns are similar for some personal liability businesses, in commercial liability insurance lines, such as workers' compensation and medical malpractice,  loss development varies a lot between companies depending on types of claims and length of litigation process. It remains an open question to analyze loss development heterogeneity and its links to insurance company characteristics.
	
	Occasionally, there are anomalies in loss development. For example, a few companies might have extremely large incremental paid losses in one of the later development years. Whether it is accounting error, data corruption or large claims that coincide to have payments in the same year, such anomaly would not represent the average loss development experience. Hence, it is crucial when making conclusions about industry-wide loss development patterns, to identify and make proper treatments to such anomalies/outliers. Loss development patterns and anomalies have a direct impact on a P\&C insurer's financials in terms of reserves and capital. For regulators overseeing the entire industry and reinsurers interested in exposures to the P\&C market, knowing the similarity and particularity of loss development can be helpful in setting strategic plans and making decisions. 
	
	The study of loss development has largely revolved around historical paid loss in the form of \emph{loss triangles}. In fact, every P\&C insurer is required to record business line level loss triangles for the past 10 accident years as part of Schedule P in their annual regulatory NAIC (National Associate of Insurance Commissioners) filings. Table~\ref{SampleTriangle} lists a sample loss triangle of cumulative paid losses with a full 10-year loss development for the accident year nine years prior to the current year, and partial loss development for the years thereafter.  	Development triangles also feature in non-insurance applications, for example for tracking disease cases that often have a delay between diagnosis and reporting, or hospital data that tracks infections and appearance of symptoms.  Loss triangles are key for actuarial reserving \citep{m93, wm08, rka16}, which involves predicting losses to be paid out in the future years, a.k.a.~the lower right part of the loss triangle. Reserving analyses use parametric or non-parametric methods based on a triangle of 55 data points. However, the limitations of such cell-based methods motivate a functional perspective, as developed in this paper.
    
In the later development lags, the scarcity of data points leads to unreliable point estimates and their standard errors. With loss triangle data available at business line level for many years, insurers can borrow information from industry data to improve its individual forecasts while accounting for heterogeneity of triangles. In our analysis below we explore runoff loss triangles in workers' compensation business line for 100+ companies in the U.S., aiming to pool multiple triangles. There is an emerging strand of literature exploring dependence structure between business lines \citep{sf11,zd13} and studying the triangles in multi-company context \citep{shi17}. We also mention the related works that deal with individual claims development, aiming to leverage policy-level covariates to predict loss development \citep{wuthrich18,taylor19,kuo2019,gabrielli21,cai24}. Recently a variety of neural network methods \citep{wuthrich18,schneider2025advancing} have been proposed.

Beyond projecting the most likely trajectory, it is important to also quantify uncertainty regarding future loss development factors. A common strategy for that purpose is bootstrapping \citep{steinmetz2024bootstrap}. Alternatively, there are probabilistic predictive models, such as Gaussian-process based methods \citep{ludkovski2020gaussian,lally2018estimating,ang2022hierarchical}, and multi-output GPs \citep{griffith2025multioutput}.  From a parametric perspective, multivariate dependence can be captured by Sarmanov distributions \citep{abdallah2023rank}, copula \citep{sf11}, categorical embedding \citep{shi2023non}, and time series techniques \citep{nieto2021gamma}.
%
%
\cite{sh16} considered a  Bayesian hierarchical model where a single hyper-prior is used for loss developments.  However, their method does not allow for grouping by covariates or any distance metrics we envision. 

From the statistical direction, we leverage functional principal component analysis techniques (PCA) and functional depth methods.
The version of functional PCA with penalized linear regression is a modification of one by \citet{shang13}; see also \cite{shang2017forecasting,Elias2022,shang2011nonparametric}. For depth methods we refer to \cite{dai2020,extremaldepth,sun2012exact}. 


\subsection{Contributions and Roadmap}
We present a study of the aggregate paid loss development in the property and casualty (P\&C) insurance market using functional data analysis tools. In contrast to 
considering a given triangle as the base object to be modeled, we adopt a \emph{functional data} viewpoint, thinking of loss development as curves,  namely as a series of Incremental Loss Ratios (ILR) or Cumulative Loss Ratios (CLR), indexed by the lag year $x$. Moreover, instead of relying on a Markov property that underlies, say the Mack Chain Ladder method and assumes that the next ILR depends just on the previous year, we allow for a full path-dependency. This concept is critical for sorting among hundreds of triangles. 

Our initial goals are (i) identifying commonalities in loss development patterns within a business line; (ii) detecting anomalies in loss development curves; (iii) investigating similarities and differences in development patterns pertaining to groups of company-specific factors (e.g.~capital structure, region) and across different historical periods. In the second half of the paper we leverage the insights from (i)-(iii) to introduce a novel (iv) probabilistic forecasting of loss development given a partially developed curve. 

Functional data depth provides a rigorous tool to identify central patterns and anomalies in the functional loss ratio data. Thus, from a practical perspective, it helps us learn the relationship between loss development curves through a single metric and to get a better picture of typical development shape, identify anomalous observations via statistical outlier techniques, and mitigate the impact of the respective outlier data. Similarly, we propose to exploit the functional viewpoint to probabilistically forecast future ILRs.  Our approach below could be labeled semi-parametric: in addition to using regression to capture similarity of different curves and employing the resulting fit to project future loss ratios, we also directly ``extend'' the existing partially developed history in a functional manner rather than cell-by-cell. Moreover, we use functional bootstrapping to quantify future uncertainty jointly across all the future lags. 

	Our findings show that the loss development pattern can be dissected down to (a) near-term (development lag 0), (b) medium-term (lags 1-2) and (c) long-term (lags 3-9) components. Most outliers occur in the long-term segment, while the most variation in patterns happens in the medium-term segment. There are also some extreme short-term patterns, but given the variability across firms, most of these should not be classified as true outliers.
    
    We then examine the aggregate patterns based on time-invariant company characteristics, including business focus (commercial vs.~personal business), geographical focus (national vs.~regional), capital structure (mutual vs.~stock), and across historical periods from accident year 1987 to 2010. We find distinctive loss development patterns between commercial and personal business. The companies focusing on the Midwest region have a different development pattern than the rest, and the late 80s' pattern differs from the 90s' and from the 2000's. This study of association between company factors and loss development, to our knowledge, does not exist in either empirical insurance or the actuarial science literature.  

    For probabilistic completion of partially developed losses, we find that ILR analysis is more flexible, in particular to remove outliers prior to fusing many curves into a single dataset. We also demonstrate that our approach provides better predictive intervals than CL (which tends to underestimate future uncertainty), while providing comparable MSE on ultimate CLR. Last but not least, we show the utility of building functional predictive envelopes beyond the standard marginal credible intervals.

    Empirically, our work lays the foundation for insurance companies to improve on loss reserving by pooling company specific information and industry-wide data. The tools are also useful for regulators to detect unusual loss development patterns and study patterns by region and company types. Our findings shed light on fusing information from industry-wide Schedule P data and on bootstrapping empirical triangles to make non-parametric probabilistic forecasts of future loss development. 
    

The rest of the paper is organized as follows. Section \ref{sec:data} presents our data, and we perform an exploratory functional analysis of our dataset that spans 137 insurance companies in Section \ref{sec:explanatory}. Section \ref{sec:forecast} describes our main proposal of functional forecasting of incomplete development curves. Moving from some illustrative results in Section \ref{sec:forecast}, Section \ref{sec:assessment} quantifies the performance of our method and benchmarks it against the classical Mack Chain-ladder (CL) approach.
Section \ref{sec:conclusion} concludes. 
	
	\section{Data}\label{sec:data}

Our empirical analysis relies on industry-wide loss development data for the U.S. workers’ compensation line of business. The data\footnote{Provided to us by a leading insurance analytics firm which undertook proprietary data wrangling and cleaning} originate from Schedule P of the National Association of Insurance Commissioners (NAIC) statutory filings, which require every property and casualty (P\&C) insurer to report annual business line loss triangles for the past ten accident years. These triangles form the foundation of regulatory reserving analyses and are widely used in academic and practitioner research (see, e.g., \citealp{m93, wm08, rka16}).

\subsection{Structure of the Data}
For each company $i$ and accident year $t$, the cumulative paid losses are recorded across development lags $x=0,\dots,9$, denoted $C_{i,t}(x)$. Incremental losses are obtained by
\[
Y_{i,t}(x) = 
\begin{cases}
C_{i,t}(x) - C_{i,t}(x-1), & x > 0, \\
C_{i,t}(0), & x = 0,
\end{cases}
\]
and normalized by the corresponding net premium earned $P_{i,t}$ to form incremental and cumulative loss ratios:
\[
y_{i,t}(x) = \frac{Y_{i,t}(x)}{P_{i,t}}, 
\qquad 
c_{i,t}(x) = \frac{C_{i,t}(x)}{P_{i,t}}.
\]
These ratios provide scale-free measures of loss development, enabling meaningful cross-company comparisons (\cite{sf11, zd13, shi17}). We analyze triangles of \emph{cumulative paid losses} as displayed in Table~\ref{SampleTriangle}, emphasizing our cross-sectional focus with 100+ companies and 3000+ company-accident year development curves. 

\begin{table}[!ht]
		\centering
		{\scriptsize 
			\begin{tabular}{cccccccccccc}
				\toprule 
				Accident & Earned & 0 & 1 & 2 & 3 & 4 & 5 & 6 & 7 & 8 & 9 \\ 
				Year & Premium & &&&&&&&&& \\ [2pt]
				\midrule
				$t_i$ & $ P_{i,t_i} $ & $ C_{i,t_i}(0) $ & $ C_{i,t_i}(1) $ & ... & ... & ... & ... & ... & ... & ... & $ C_{i,t_i}(9) $ \\ [2pt]
                ... & ... & ... & ... & ... & ... & ... & ... & ... & ... & ... & ... \\ [2pt]
                $T-9$ & $ P_{i,T-9} $ & $ C_{i,T-9}(0) $ & $ C_{i,T-9}(1) $ & ... & ... & ... & ... & ... & ... & ... & $ C_{i,T-9}(9) $ \\ [2pt]
				$T-8$ & $ P_{i,T-8} $ & $ C_{i,T-8}(0) $ & ... & ... & ... & ... & ... & ... & ... & $ C_{i,T-8}(8) $ &  \\  [2pt]
				$T-7$ & $ P_{i,T-7} $ & $ C_{i,T-7}(0) $ & ... & ... & ... & ... & ... & ... & ... &  &  \\  [2pt]
				$T-6$ & $ P_{i,T-6} $ & $ C_{i,T-6}(0) $ & ... & ... & ... & ... & ... & ... &  &  &  \\ [2pt]
				$T-5$ & $ P_{i,T-5} $ & $ C_{i,T-5}(0) $ & ... & ... & ... & ... & ... &  &  &  &  \\ [2pt]
				$T-4$ & $ P_{i,T-4} $ & $ C_{i,T-4}(0) $ & ... & ... & ... & ... &  &  &  &  &  \\ [2pt]
				$T-3$ & $ P_{i,T-3} $ & $ C_{i,T-3}(0) $ & ... & ... & ... &  &  &  &  &  &  \\ [2pt]
				$T-2$ & $ P_{i,T-2} $ & $ C_{i,T-2}(0) $ & ... & $ C_{i,T-2}(2) $ &  &  &  &  &  &  &  \\ [2pt]
				$T-1$ & $ P_{i,T-1} $ & $ C_{i,T-1}(0) $ & $ C_{i,T-1}(1) $ &  &  &  &  &  &  &  &  \\ [2pt]
				$T$ & $ P_{i,T} $ & $ C_{i,T}(0) $ &  &  &  &  &  &  &  &  &  \\
				\bottomrule 
		\end{tabular}}
		\caption{Sample cumulative paid losses triangle in calendar year $T$ for one line in one company.} 
		\label{SampleTriangle}
	\end{table}

We next describe how we selected the final dataset from these raw filings.

\subsection{Sample Selection}
We focus on workers’ compensation triangles from 1987 to 2010. After preprocessing, the dataset comprises 3,362 company-year observations across more than 100 insurers, each with at least \$1 million in net premium earned per accident year. This threshold excludes marginal portfolios with negligible exposure, thereby ensuring the stability of loss ratio curves. To reduce distortions from extraordinary corporate events, we additionally exclude companies whose earned premium varies by more than a factor of 10 across accident years, since such swings typically reflect acquisitions, divestitures, or book transfers that break comparability of historical development.

Each insurer is identified by an NAIC code (e.g., ``C121’’), and we enrich the triangle data with time-invariant company characteristics:  
(i) \emph{Geographic focus} (National, Midwest, Northeast, South, West);  
(ii) \emph{Ownership structure} (Stock, Mutual, Other);  
(iii) \emph{Business focus} (Personal, Commercial, Workers’ Compensation).  
(those covariate are {\em time-invariant}, as opposed to earned premiums, that is inherently {\em time-varying}).
This augmentation allows us to investigate heterogeneity in development patterns across company types, consistent with earlier multi-company reserving studies (e.g., \citealp{shi17, gabrielli21}).

\subsection{Descriptive Features}
\begin{figure}[!ht]
		\centering
		\includegraphics[width=0.9\textwidth]{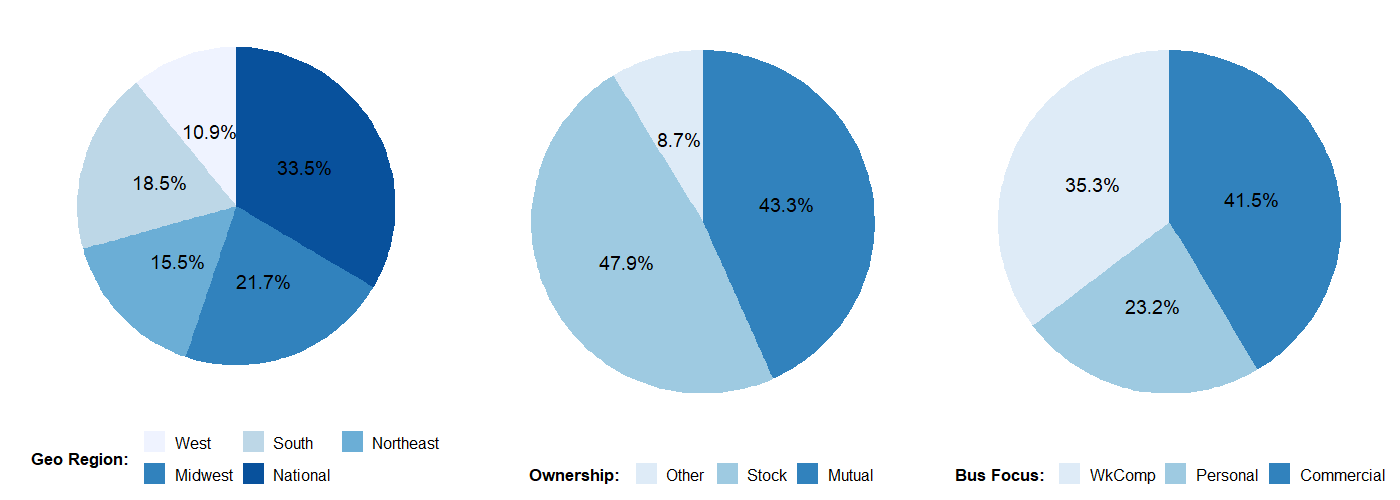}
		\caption{Distribution of sample development triangles by Geographical Region, Ownership Structure and Business Focus.}
		\label{fig:piecharts}
	\end{figure}

Figure~\ref{fig:piecharts} summarizes the distribution of observations by region, ownership, and business focus. Across accident years, the number of reporting insurers varies between 70 and 224, with 115 companies contributing at least 15 years of observations. This breadth of coverage provides a rich empirical basis for functional data analysis of loss development. In contrast to traditional single-triangle reserving studies, our cross-sectional perspective allows pooling of information across insurers while accounting for systematic differences in underwriting portfolios and corporate structure.
	
\bigskip

In sum, our dataset combines the regulatory completeness of NAIC Schedule P triangles with additional company-level covariates, thereby enabling us to study not only typical loss development patterns but also the heterogeneity and anomalies that arise across insurers and time. This design sets the stage for the functional data methods developed in the following sections.

\section{Exploratory Analysis of Runoff Triangles}\label{sec:explanatory}
    
Before developing our forecasting framework, we first investigate the empirical properties of incremental loss ratio curves in our dataset. Exploratory analysis is essential for understanding both the typical shape of loss development and the anomalies that may distort statistical modeling. This section provides descriptive statistics, discusses outlier detection, and examines heterogeneity across company characteristics.

\subsection{Descriptive Statistics}

Table~\ref{SummaryILR} reports summary statistics of ILRs across development lags $x=0,\dots,9$ for workers’ compensation paid loss triangles from 1987--2010. The non-robust mean and standard deviation at each lag are systematically larger than the robust measures (median and median absolute deviation, MAD, see \cite{rc93}). This discrepancy highlights the presence of outlying curves, consistent with earlier findings in multi-company analyses (\citealp{shi17, gabrielli21}). 

The ILR distribution exhibits a characteristic shape: losses peak at lag 1, then decline sharply at lags 2--3, followed by a long right tail extending up to lag 9. Beyond lag 7, mean ILRs fall below 1\%, reflecting the long-tailed settlement patterns in workers’ compensation claims (\citealp{taylor19}). Negative ILRs, corresponding to downward adjustments in cumulative paid losses, occur in later lags and introduce skewness and heavy tails.

\begin{table}[!ht]
		\centering
		{\scriptsize 
			\begin{tabular}{lrrrrrrrrrrr}
				\toprule
				Lag $\quad x$ & \multicolumn{1}{c}{0} & \multicolumn{1}{c}{1} & \multicolumn{1}{c}{2} & \multicolumn{1}{c}{3} & \multicolumn{1}{c}{4} & \multicolumn{1}{c}{5} & \multicolumn{1}{c}{6} & \multicolumn{1}{c}{7} & \multicolumn{1}{c}{8} & \multicolumn{1}{c}{9}\\
				\midrule
				\rowcolor{gray!6} Mean & 0.1669 & 0.1900 & 0.1003 & 0.0581 & 0.0344 & 0.0219 & 0.0145 & 0.0109 & 0.0078 & 0.0062\\
				\rowcolor{gray!6} Std. Dev. & 0.0598 & 0.0636 & 0.0429 & 0.0342 & 0.0256 & 0.0212 & 0.0184 & 0.0187 & 0.0166 & 0.0146\\
				\rowcolor{gray!6} Median & 0.1628 & 0.1843 & 0.0948 & 0.0547 & 0.0323 & 0.0196 & 0.0128 & 0.0086 & 0.0062 & 0.0047\\
				\rowcolor{gray!6} MAD & 0.0476 & 0.0556 & 0.0340 & 0.0249 & 0.0184 & 0.0137 & 0.0108 & 0.0084 & 0.0068 & 0.0054\\
				\addlinespace
				Min & 0.0008 & -0.1038 & -0.1677 & -0.3966 & -0.1881 & -0.2891 & -0.2785 & -0.2808 & -0.3155 & -0.2311\\
				Max & 0.6936 & 0.6647 & 0.4961 & 0.3981 & 0.3643 & 0.3492 & 0.2958 & 0.3224 & 0.2697 & 0.3293\\
				95\% quantile & 0.0817 & 0.1025 & 0.0419 & 0.0150 & 0.0039 & 0.0009 & 0.0000 & 0.0000 & -0.0002 & -0.0001\\
				99\% quantile & 0.2639 & 0.2988 & 0.1773 & 0.1121 & 0.0726 & 0.0510 & 0.0378 & 0.0297 & 0.0234 & 0.0192\\
				Skewness & 1.4431 & 0.7565 & 0.9805 & 0.7105 & 1.0798 & 0.6508 & -0.0472 & 3.5355 & -2.9519 & 2.6788\\
                Kurtosis & 7.84 & 3.19 & 5.12 & 21.72 & 19.34 & 51.10 & 61.89 & 110.70 & 137.08 & 139.89 \\
				5\% winsorized mean & 0.1651 & 0.1891 & 0.0995 & 0.0572 & 0.0338 & 0.0214 & 0.0144 & 0.0104 & 0.0077 & 0.0060\\
				5\% winsorized sd & 0.0479 & 0.0532 & 0.0354 & 0.0253 & 0.0182 & 0.0135 & 0.0105 & 0.0084 & 0.0067 & 0.0055\\
				\addlinespace
				\rowcolor{gray!6} \# Positive & 3362 & 3358 & 3355 & 3328 & 3290 & 3240 & 3173 & 3114 & 3051 & 3015\\
				\rowcolor{gray!6} \# Zero & 0 & 0 & 0 & 2 & 10 & 24 & 47 & 82 & 133 & 171\\
				\rowcolor{gray!6} \# Negative & 0 & 4 & 7 & 32 & 62 & 98 & 142 & 166 & 178 & 176\\
				\bottomrule
		\end{tabular}}
		\caption{Summary statistics of ILRs for workers' compensation paid loss triangles from 1987 to 2010}
		\label{SummaryILR}
	\end{table} 

\subsection{Outlier Detection}\label{sec:outlier-detection}

Outliers in development curves may arise from data errors, reinsurance events, or extraordinary claims. Their presence can substantially bias traditional reserving techniques (\citealp{wuthrich18}). To address this, we adopt two complementary approaches: robust principal component analysis (RPCA, \citealp{cfo07pcagrid}) and functional data depth methods (\citealp{sun2012exact, extremaldepth, dai2020}).

\begin{figure}[!htb]
		\centering
		\begin{subfigure}[b]{0.45\textwidth}
			\includegraphics[width=\textwidth]{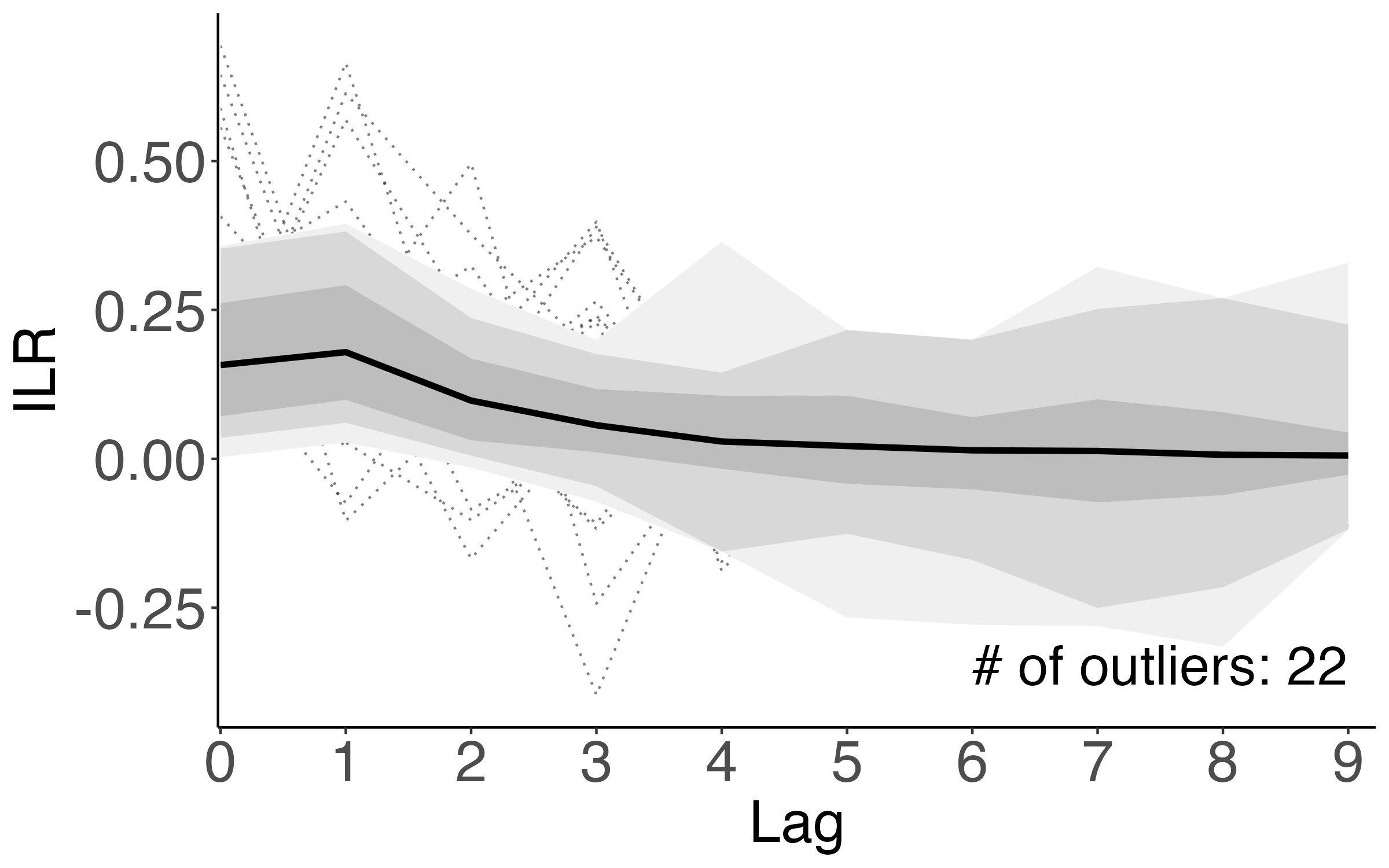}
			\caption{ {\footnotesize Extremal Depth (EXD) on first 5 RPCA scores}}
			\label{outliers_rpca}
		\end{subfigure} 
        \begin{subfigure}[b]{0.45\textwidth}
			\includegraphics[width=\textwidth]{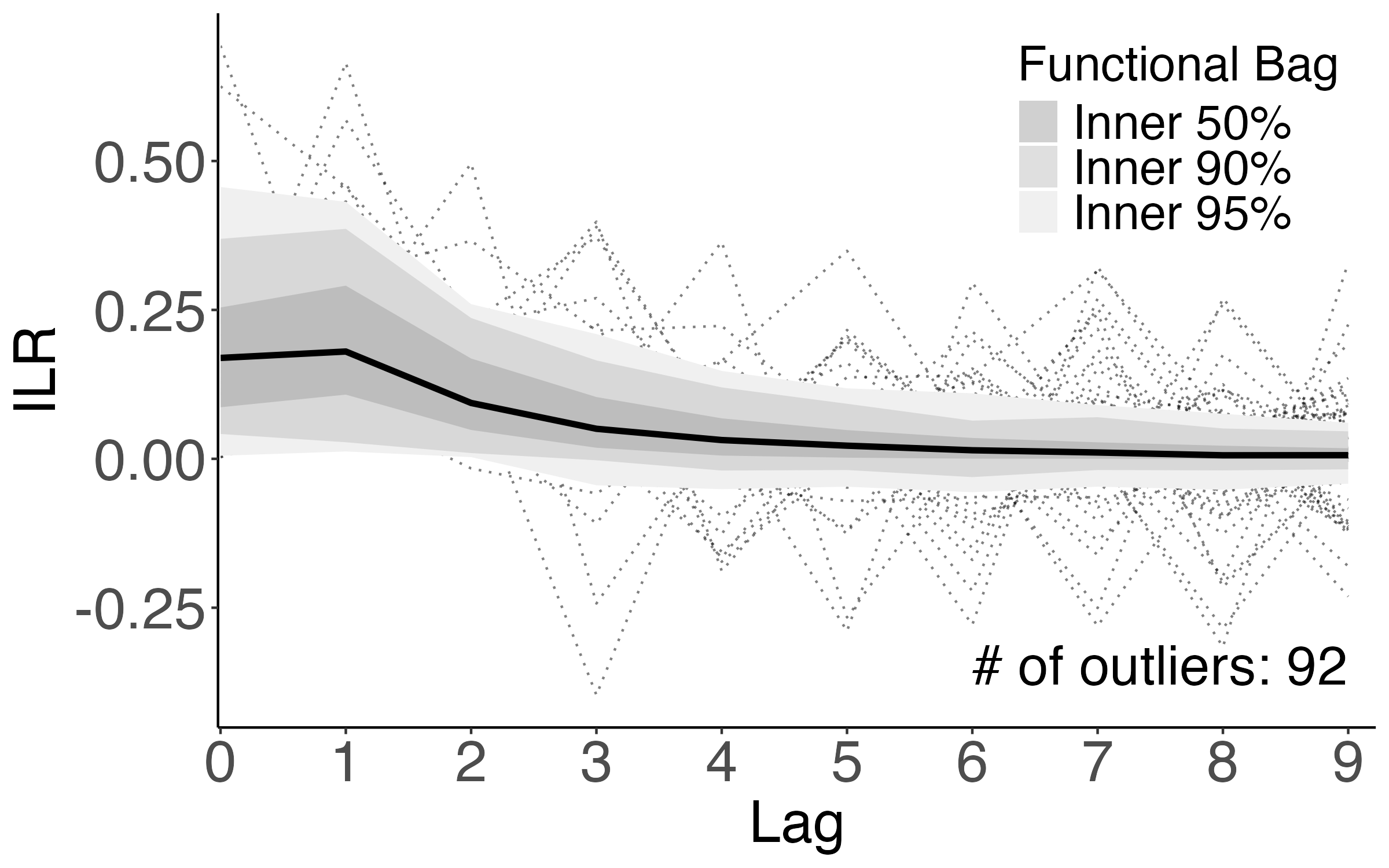}
		\caption{{\footnotesize Extremal Depth (EXD)}} \label{outliers_exd}
		\end{subfigure}\\
        
		\begin{subfigure}[b]{0.45\textwidth}
			\includegraphics[width=\textwidth]{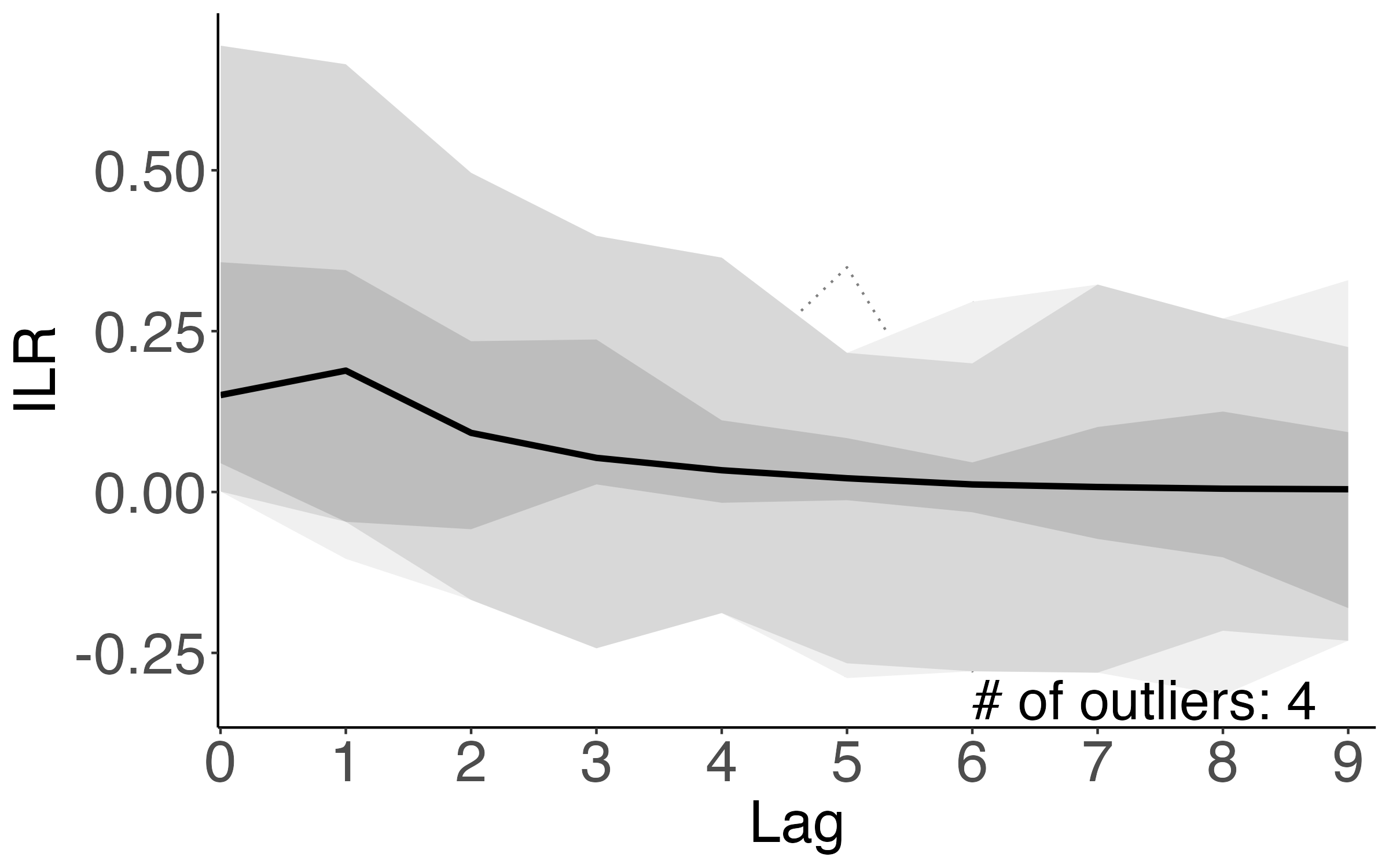}
			\caption{{\footnotesize Modified Bandwidth Depth (MBD)}}
			\label{outliers_mbd}
		\end{subfigure} 
        \begin{subfigure}[b]{0.45\textwidth}
			\includegraphics[width=\textwidth]{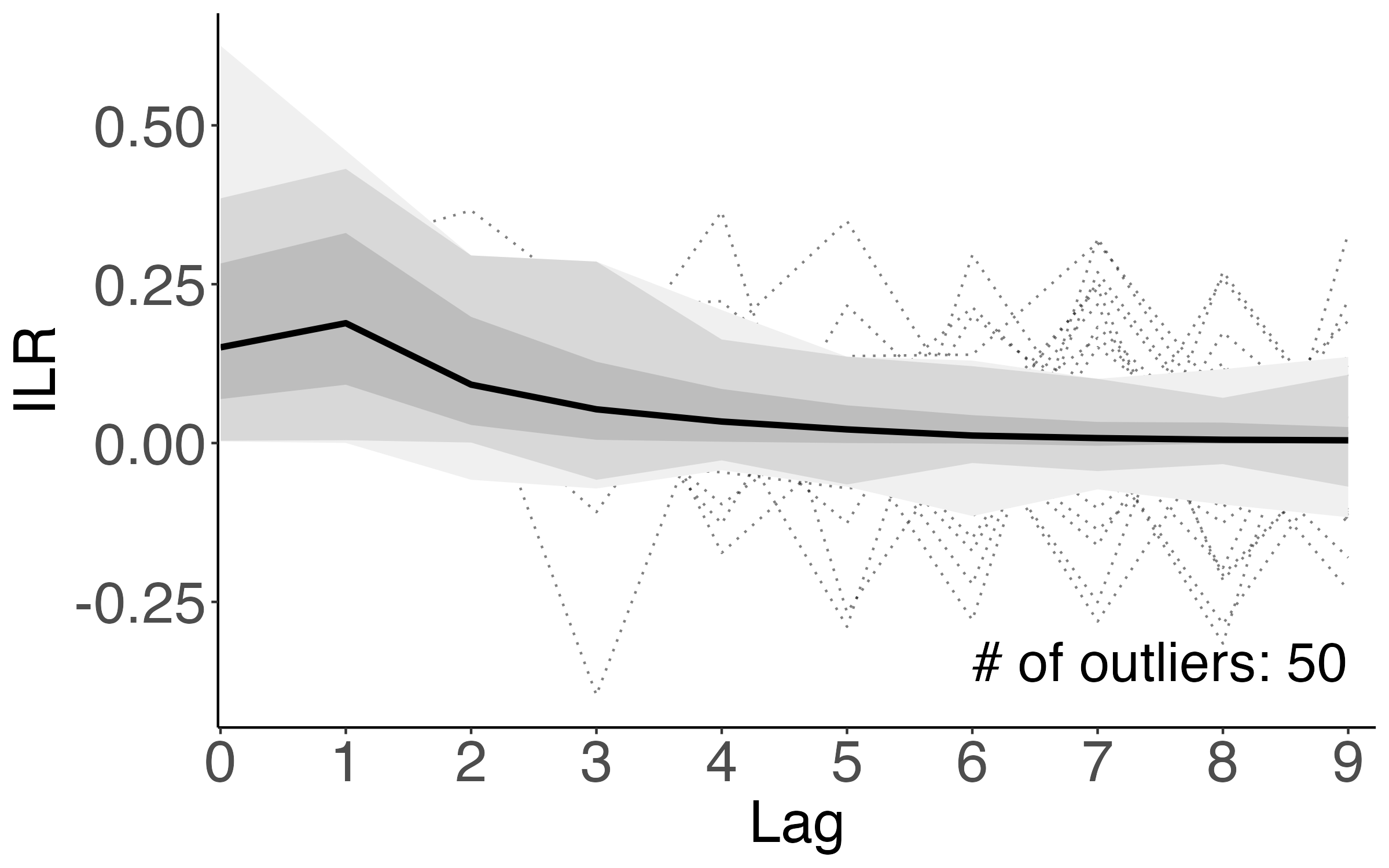}
		\caption{{\footnotesize Bandwidth Depth (BD)}} \label{outliers_bd}
		\end{subfigure}\\

		\caption{Identification of outlying loss development curves. We compare three different functional depth metrics to identify extreme ILR curves.}
		\label{fig:fdaoutlier}	
	\end{figure}

RPCA identifies curves with extreme component scores relative to the cross-sectional distribution. While effective at detecting early-lag anomalies, RPCA tends to produce overly wide tolerance bands at longer lags, limiting its discriminating power. In contrast, functional depth provides an ordering of curves from central to extreme, yielding interpretable notions of functional medians and quantile envelopes. We consider three widely used depth notions: Band Depth (BD), Modified Band Depth (MBD), and Extremal Depth (EXD). Consistent with the recommendations in \citet{extremaldepth}, we find EXD to be most effective, producing tight central envelopes and flagging curves with unusual long-tail behavior (Figure~\ref{fig:fdaoutlier}	).

The number of curves classified as outliers varies significantly by method: MBD flags only a handful of series, BD captures approximately 50 curves, while EXD identifies over 90. This sensitivity underscores the importance of choosing an outlier detection method consistent with actuarial intuition: wide variability in short lags but tighter envelopes in later lags where reserves are most sensitive. Additional technical details (with proper definitions of BD, MBD, and EXD) are provided in Appendix~\ref{app:outliers}.

\subsection{Impact of Covariates}

Having identified central patterns and outliers, we next investigate systematic heterogeneity driven by insurer characteristics.
Beyond industry-wide averages, loss development is influenced by systematic company- 
and market-level characteristics. Understanding these covariates is crucial both for 
regulators, who monitor systemic risks across heterogeneous insurers, and for actuaries, 
who must ensure that reserving models remain calibrated to the specific portfolio under study. 
In this subsection, we stratify ILR curves by several company features 
and discuss the resulting heterogeneity.

\begin{figure}[!ht]
		\centering
		\begin{subfigure}[b]{0.45\textwidth}
			\includegraphics[width=\textwidth]{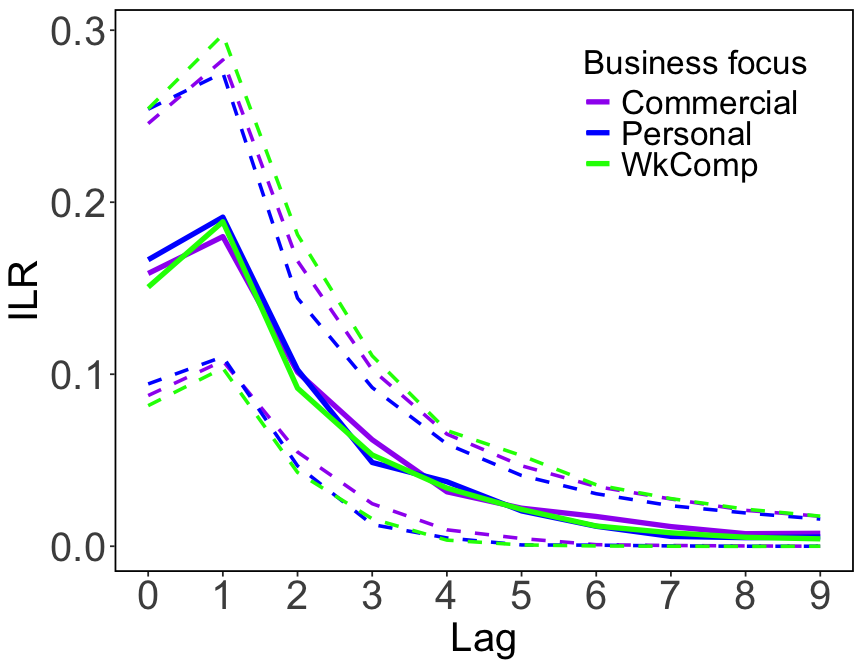}
            \caption{Business focus}
            \label{plot_bf}
		\end{subfigure}	
		\begin{subfigure}[b]{0.45\textwidth}
			\includegraphics[width=\textwidth]{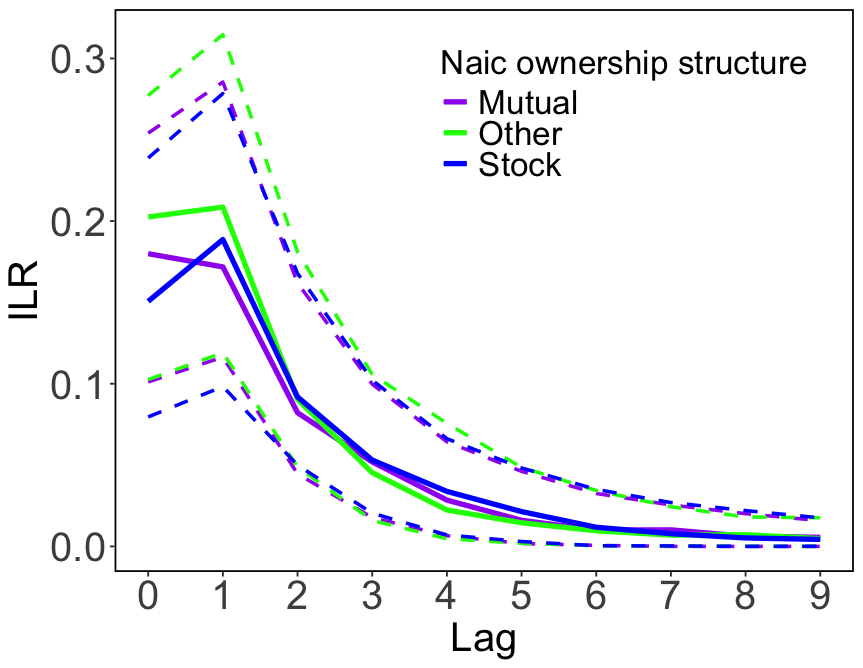}
            \caption{Ownership structure}
            \label{plot_nos}
		\end{subfigure}
		
		\begin{subfigure}[b]{0.45\textwidth}
			\includegraphics[width=\textwidth]{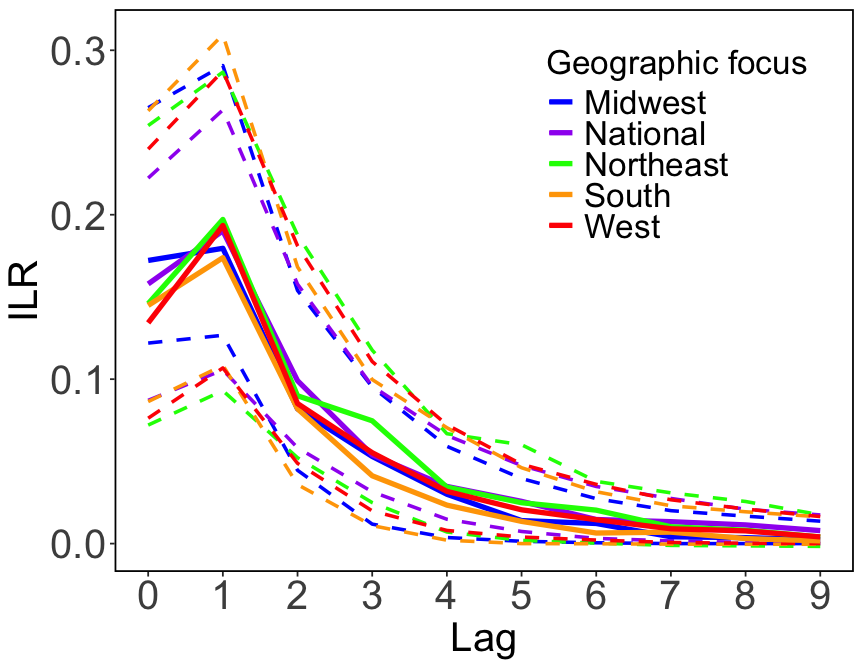}
            \caption{Geographic region}
            \label{plot_gf}
		\end{subfigure}	
		\begin{subfigure}[b]{0.45\textwidth}
			\includegraphics[width=\textwidth]{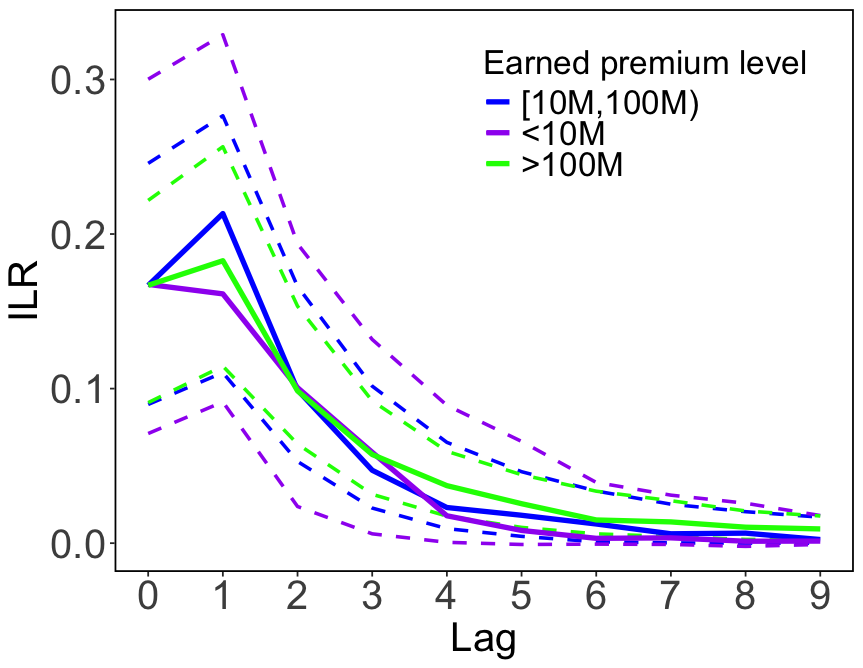}
            \caption{Earned premium level}
            \label{plot_epl}
		\end{subfigure}
    \caption{Median ILR curves and EXD-based 50\% envelopes stratified by company characteristics: (a) business focus, (b) ownership, (c) geographic region, (d) earned premium category.}
\label{fig:covariates}
    \end{figure}

\paragraph{Business focus.} 
We first distinguish insurers according to their primary line of business: 
personal, commercial, and workers’ compensation specialists. Figure~\ref{fig:covariates}(a) 
shows that median ILR curves are broadly similar across groups, with overlapping 
EXD-based 50\% envelopes. This suggests that, within the workers’ compensation line, 
loss development is relatively stable across broader business strategies. 
This finding echoes the results of \citet{shi17}, who documented strong within-line 
consistency of development patterns even for diversified insurers. 
Nevertheless, slight differences at lag 0 and lag 1 may reflect different claims handling 
speeds across personal versus commercial insurers.

\paragraph{Ownership structure.}
Ownership type (stock, mutual, or other) can affect corporate governance, 
risk appetite, and reserving practices (\citealp{CumminsWeiss2000}). 
In our dataset, Figure~\ref{fig:covariates}(b) shows modest differences: 
mutual companies tend to report slightly higher ILRs in early lags, 
potentially reflecting more conservative settlement practices, 
whereas stock companies display marginally faster runoff. 
Although the median differences are small, this dimension remains important for 
reserve adequacy studies, since ownership type may interact with capital management 
and regulatory scrutiny.

\paragraph{Geographic focus.}
Geographic diversification is another relevant factor. 
Regional insurers (Midwest, Northeast, South, West) may be more exposed to 
jurisdictional differences in workers’ compensation law, 
medical inflation, or litigation practices (\citealp{Weiss2001}). 
As shown in Figure~\ref{fig:covariates}(c), median curves exhibit mild regional variation, 
with Midwest-focused insurers displaying slightly higher lag-3 and lag-4 ILRs. 
Such effects, although subtle in aggregate, can accumulate to material differences 
in ultimate loss ratios, and highlight the importance of incorporating jurisdictional 
heterogeneity into multi-company reserving analyses.

\paragraph{Earned premium (company size).}
Company size, proxied by annual net premium earned (NPE), reveals a more striking 
pattern (Figure~\ref{fig:covariates}(d)). We partition firms into three groups: 
small ($<$\$10m annual NPE), medium (\$10m–\$100m), and large ($>$\$100m). 
Small insurers display the greatest volatility, with wider envelopes across nearly all lags, 
consistent with classical credibility theory (\citealp{Buhlmann1967}). 
Medium-sized insurers exhibit the sharpest peak at lag 1, followed by steep declines, 
while large insurers show smoother curves and persistently higher ILRs beyond lag 4. 
This suggests that larger insurers, possibly due to more complex claims portfolios 
and protracted litigation, settle a higher proportion of losses at longer horizons. 
The empirical evidence thus aligns with practitioner intuition: 
smaller companies tend to close claims more quickly but with greater variability, 
whereas larger companies carry longer tails but exhibit more stable patterns.

\paragraph{Calendar time.}
Finally, we investigate accident-year effects, which capture market-wide cycles 
in workers’ compensation. Figure~\ref{fig:covariates}(e) reveals recurrent peaks in median 
ILRs during the late 1980s, mid-1990s, and late 2000s. 
These coincide with well-documented underwriting cycles 
(\citealp{CumminsDohertyLo1992, HarringtonNiehaus2000}) 
and suggest that external factors—such as interest rate environments, 
medical cost inflation, and regulatory changes—play a significant role in shaping 
development patterns. Notably, while levels fluctuate across periods, 
we do not observe a monotone trend in ILRs, underscoring the cyclical rather than 
secular nature of these effects.

Taken together, these results highlight three key insights. 
First, ownership and business focus appear to exert only limited influence, 
consistent with the dominance of line-specific settlement dynamics. 
Second, company size is a primary driver of heterogeneity, with small insurers 
displaying the most volatile development. 
Third, accident-year effects reveal cyclical variation linked to market conditions. 
These findings reinforce the need for models that can flexibly account for both 
cross-sectional and temporal heterogeneity, motivating our functional approach 
in Section~\ref{sec:forecast}.

    \begin{figure}[!ht]    
		\label{middlecs}
        \centering
		\begin{subfigure}[b]{0.475\textwidth}
			\includegraphics[width=\textwidth]{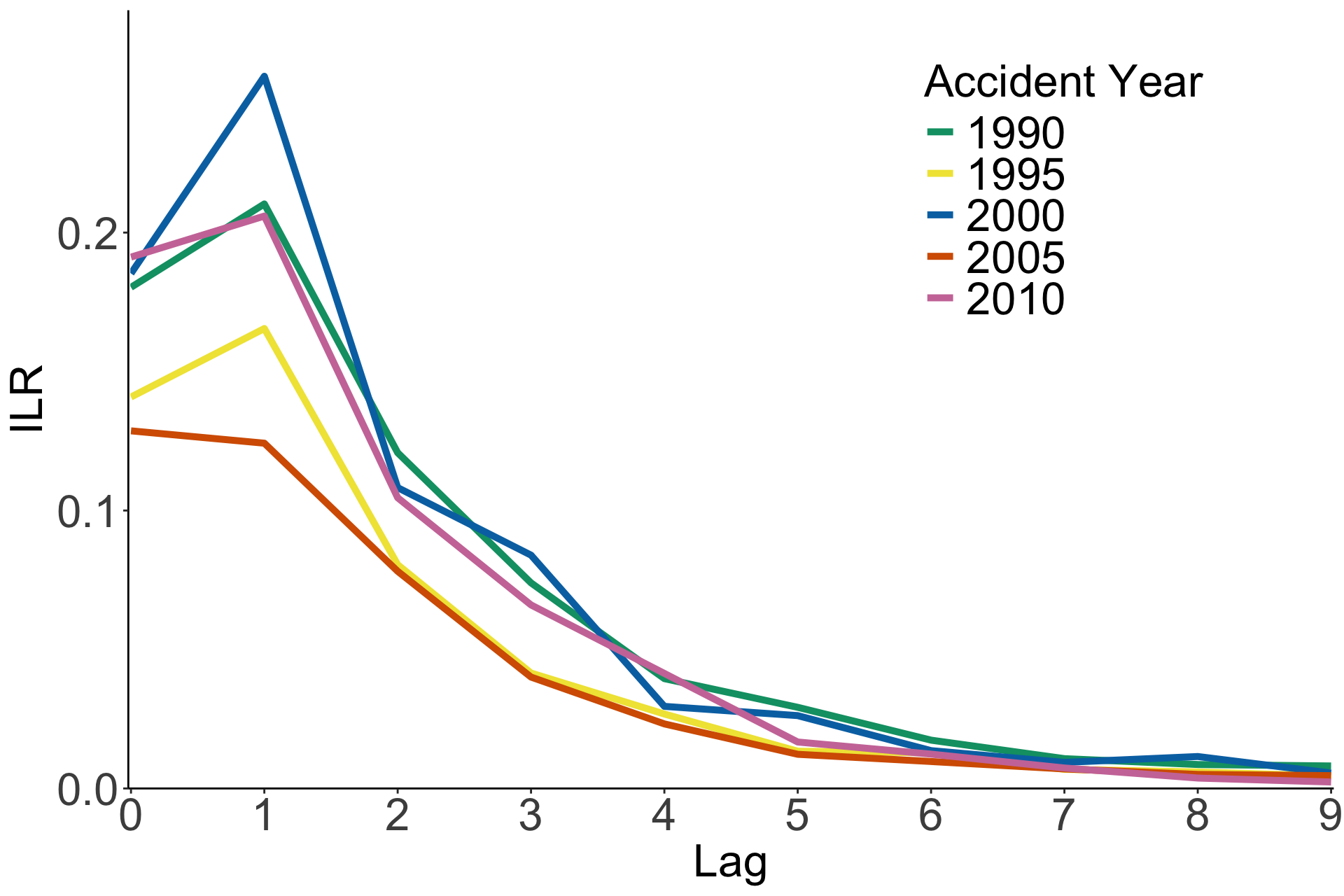}
            \caption{ILR median curve for select AYs}
            \label{plot_ay_median}
		\end{subfigure}
		\begin{subfigure}[b]{0.475\textwidth}
			\includegraphics[width=\textwidth]{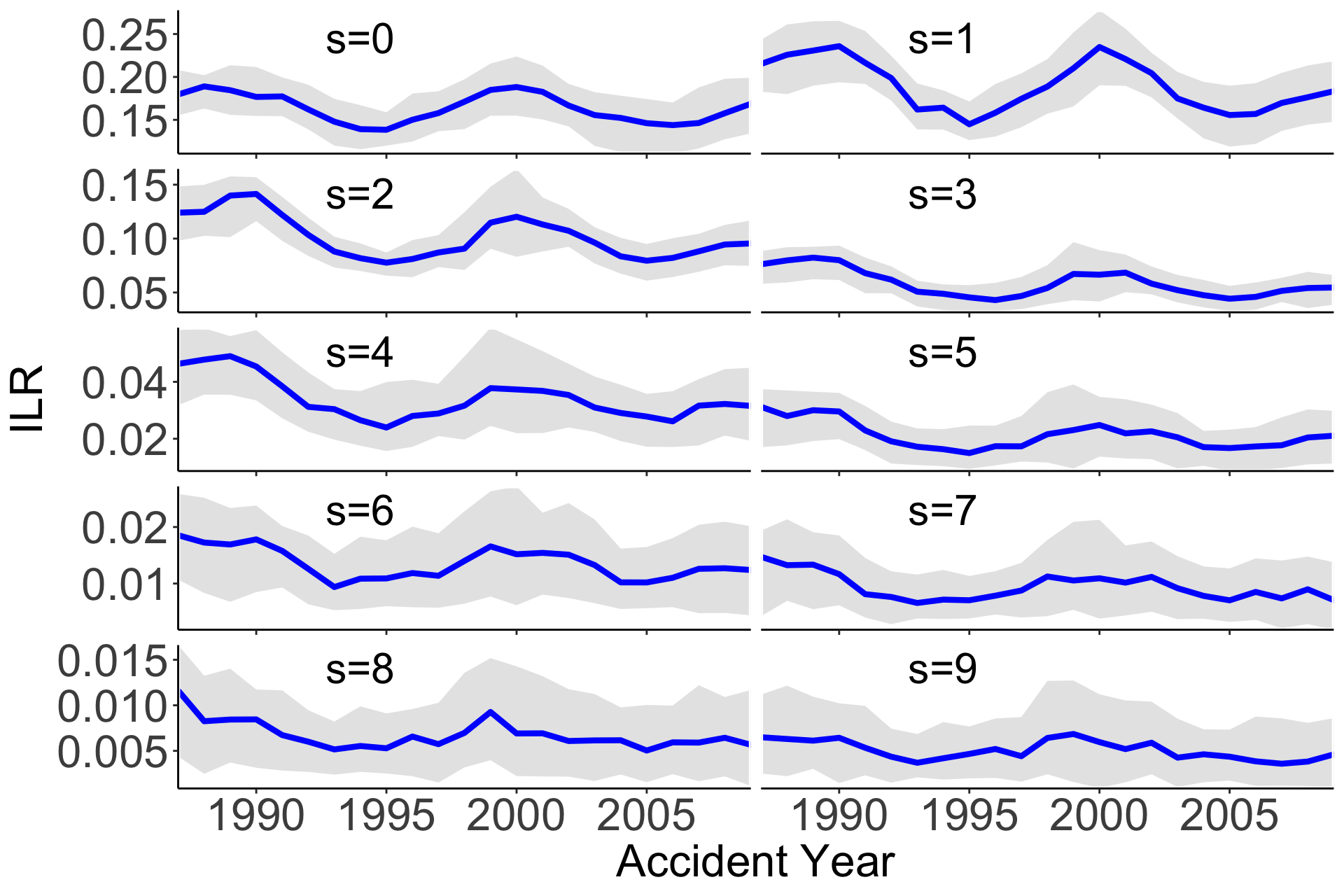}
            \caption{Median and interquartile range of ILR by lag $s$. $y$-axes vary by row. }
            \label{plot_ay_by_lag}
		\end{subfigure}
		\label{plot_ay}
        \caption{ILR in relation to accident year}
	\end{figure}

\section{Forecasting Incremental Loss Ratios (ILR) Curves}\label{sec:forecast}

We now develop a probabilistic forecasting framework for ILR curves 
based on partially observed loss development triangles. 
Our aim is to forecast the continuation of a curve $y_{n^\ast}(x)$, 
observed only up to lag $s-1$, by exploiting information contained in the historical database. 
The methodology builds on functional principal component analysis (FPCA), 
regression of functional scores on covariates, and penalized curve completion. 
We also construct bootstrap-based predictive intervals to quantify uncertainty.

\subsection{Functional Representation and Matrix Formulation}
Let $\mathcal{X}=\{0,\dots,9\}$ denote the set of development lags. 
For each company--accident year pair $n\in\mathcal{N} = \{i = 1,2,\ldots,I\} \times \{t = t_i, t_i+1, \ldots,T-9\}$ , 
we observe a complete ILR curve $\{y_{n}(x):x\in\mathcal{X}\}$. 
We view $y_{n}$ as a function in the Hilbert space $L^{2}(\mathcal{X})$ 
with inner product $\langle f,g\rangle=\sum_{x\in\mathcal{X}} f(x)g(x)$. 
Let $\mu(x)=\mathbb{E}[y_{n}(x)]$ denote the mean function 
and $\Gamma(x,x')=\mathrm{Cov}(y_{n}(x),y_{n}(x'))$ the covariance kernel.

By the Karhunen–Loève decomposition (see \cite{ramsay2005functional}), 
\begin{equation}
y_{n}(x) = \mu(x) + \sum_{k=1}^{\infty}\phi_{k}(x)\beta_{k,n}, 
\qquad 
\beta_{k,n} = \langle y_{n}-\mu, \phi_{k}\rangle,
\end{equation}
where $\{\phi_{k}\}$ are orthonormal eigenfunctions of $\Gamma$ 
with decreasing eigenvalues $\lambda_{1}\geq \lambda_{2}\geq \cdots \geq 0$. 
In practice, we truncate to the first $K$ components, consistently with Equation (\ref{eq:fpca}),
\begin{equation}
y_{n}(x)= \mu(x) + \sum_{k=1}^{K}\phi_{k}(x)\beta_{k,n}+\varepsilon_n(x) \approx \mu(x) + \sum_{k=1}^{K}\phi_{k}(x)\beta_{k,n}, 
\label{eq:fpca_trunc}
\end{equation}
where $K$ is chosen via cross-validation or variance explained criterion 
(\citealp{shang2011nonparametric, Elias2022}). 
The residual term represents high-order variation not captured by the factor structure.

Let $Y\in\mathbb{R}^{N\times |\mathcal{X}|}$ be the data matrix with entries $Y_{n,x}=y_{n}(x)$. 
Let $\Phi\in\mathbb{R}^{|\mathcal{X}|\times K}$ denote the matrix of estimated eigenfunctions 
$\hat{\phi}_{k}$ evaluated on $\mathcal{X}$. 
Then (\ref{eq:fpca_trunc}) may be written as
\begin{equation}\label{eq:9}
Y \approx \mathbf{1}\mu^{\top} + B\Phi^{\top},
\end{equation}
where $B\in\mathbb{R}^{N\times K}$ contains the scores $\hat{\beta}_{k,n}$. 
Thus FPCA is equivalent to low-rank factorization of $Y$.

\subsection{Regression Priors and Partial Curve Completion}

For a mathematical formulation, the idea is that each
complete ILR curve is represented through functional PCA as
\begin{equation}
y_{n}(x) = \hat{\mu}(x) + \sum_{k=1}^{K} \hat{\phi}_{k}(x) \, \hat{\beta}_{k,n} + \varepsilon_{n}(x), 
\qquad x \in \{0,\dots,9\},
\label{eq:fpca}
\end{equation}
where $\hat{\mu}$ is the empirical mean function, $\hat{\phi}_{k}$ are the estimated functional
principal components, and $\hat{\beta}_{k,n}$ are the scores for curve $n$.

Each score vector $\beta_{n}=(\beta_{1,n},\dots,\beta_{K,n})^{\top}$ 
encodes the shape of curve $n$. 
To incorporate company features $X_{n}$ (earned premium, ownership, region, etc.), 
we model
\begin{equation}
\beta_{k,n} = f_{k}(X_{n}) + \xi_{k,n}, \qquad k=1,\dots,K,
\label{eq:regression_scores}
\end{equation}
where $f_{k}$ is approximated by a linear form 
$f_{k}(X)=X^{\top}\theta_{k}$ estimated via LASSO 
(\citealp{Tibshirani1996}) to encourage sparsity. 
Let $\hat{\beta}^{\mathrm{RM}}_{k,n^\ast}=X_{n^\ast}^{\top}\hat{\theta}_{k}$ 
denote the regression-based prior score for a new curve $n^\ast$.

Suppose curve $n^\ast$ is observed up to lag $s-1$, i.e., 
we observe $y_{n^\ast}(x)$ for $x\in \xe=\{0,\dots,s-1\}$.
We complete the curve by solving a penalized least squares problem:
\begin{equation}
\hat{\beta}^{\mathrm{PLS}}_{n^\ast} 
= \argmin_{\beta\in\mathbb{R}^{K(s)}}
\Bigg\{
\sum_{x\in \xe}
\big(y_{n^\ast}(x) - \hat{\mu}(x) - \sum_{k=1}^{K(s)}\hat{\phi}_{k}(x)\beta_{k}\big)^{2}
+ \lambda(s)\sum_{k=1}^{K(s)}\big(\beta_{k}-\hat{\beta}^{\mathrm{RM}}_{k,n^\ast}\big)^{2}
\Bigg\}.
\label{eq:pls_matrix}
\end{equation}
The first term enforces fit to the partial curve, while the second shrinks towards the regression prior.

\paragraph{Closed form.}
Let $y_{e}$ be the vector of observed lags, 
$\Phi_{e}\in\mathbb{R}^{s\times K}$ the restriction of $\Phi$, 
and $\mu_{e}$ the restricted mean. Then (\ref{eq:pls_matrix}) has solution
\begin{equation}
\hat{\beta}^{\mathrm{PLS}}_{n^\ast} = 
\big(\Phi_{e}^{\top}\Phi_{e}+\lambda(s)I_{K(s)}\big)^{-1}
\big(\Phi_{e}^{\top}(y_{e}-\mu_{e})+\lambda(s)\hat{\beta}^{\mathrm{RM}}_{n^\ast}\big).
\label{eq:closed_form}
\end{equation}
This highlights the shrinkage effect of $\lambda(s)$: 
for small $s$, regression priors dominate, while as $s$ increases, 
the fit to partial observations becomes more important.

\subsection{Forecasting and Prediction Intervals}\label{sec:boostrap}
The predicted ILR curve for future lags $\xl=\{s,\dots,9\}$ is
\begin{equation}
\hat{y}_{n^\ast}(x) = \hat{\mu}(x)+\sum_{k=1}^{K(s)}\hat{\phi}_{k}(x)\,
\hat{\beta}^{\mathrm{PLS}}_{k,n^\ast}, \qquad x\in \xl.
\end{equation}
The corresponding cumulative loss ratio (CLR) forecast is
\begin{equation}
\hat{c}_{n^\ast}(x) = c_{n^\ast}(s-1)+\sum_{j=s}^{x}\hat{y}_{n^\ast}(j), 
\qquad x\in \xl.
\end{equation}

To quantify forecast uncertainty, we construct bootstrap prediction intervals. 
Let $\epsilon_{n}(x)=y_{n}(x)-\hat{\mu}(x)-\sum_{k=1}^{K(s)}\hat{\phi}_{k}(x)\hat{\beta}_{k,n}$ 
denote residuals. We assume $\epsilon_{n}$ are independent across $n$ with mean zero. 

\paragraph{Algorithm.} For $b=1,\dots,B$:  
\begin{enumerate}
\item Resample company features and curves $(X_{n},y_{n})$.  
\item Refit regressions (\ref{eq:regression_scores}), obtain $\hat{\beta}^{\mathrm{RM},(b)}$.  
\item Solve (\ref{eq:closed_form}) to obtain $\hat{\beta}^{\mathrm{PLS},(b)}_{n^\ast}$.  
\item Resample residual function $\epsilon^{(b)}$ and set
\begin{equation}
\hat{y}^{(b)}_{n^\ast}(x) = \hat{\mu}(x) + \sum_{k=1}^{K(s)}\hat{\phi}_{k}(x)\,
\hat{\beta}^{\mathrm{PLS},(b)}_{k,n^\ast} + \epsilon^{(b)}(x).
\end{equation}
\end{enumerate}
Pointwise $(1-\alpha)$ prediction intervals are obtained from the empirical quantiles 
of $\{\hat{y}^{(b)}_{n^\ast}(x)\}$. Simultaneous bands are formed by functional depth 
(\cite{extremaldepth,dai2020}), defining the central $100(1-\alpha)\%$ region. More precisely, from our bootstrapped ILR curves, we construct $100(1-\alpha)$\% central region using several functional depth methods. Let $\xe=\{0,\dots,s-1\}$ be observed lags and 
$\xl = \{s,\dots,9\}$ the missing lags. The construction follows equations (\ref{eq:fpca_trunc}) and (\ref{eq:regression_scores}) in \cite{extremaldepth}:
    \begin{equation}\label{eq:depth-region}
        C_{1 - \alpha} = \{ y \in S: y_{n}^{L}(x) \leq y(x) \leq y_{n}^{U}(x),\; \forall x \in \xl \},
    \end{equation}
    where $S = \{\hat{y}^{PLS,1}_{n^*}(x), \ldots,\hat{y}^{PLS,B}_{n^*}(x)\}, x \in \xl$ is the collection of all bootstrapped predicted ILR curves and 
    \begin{align*}
    \hat{y}_{n^*}^{D,L}(x) &:= \inf \left\{y_{n^*}(x) : y \in S,\; D(y_{n^*},\mathbb{P}_{n^*}) > \alpha\right\}, \\
    \hat{y}_{n^*}^{D,U}(x) &:= \sup \left\{y_{n^*}(x) : y \in S,\; D(y_{n^*},\mathbb{P}_{n^*}) > \alpha\right\},
    \end{align*}
    with $D$ representing a functional depth measure, and $\mathbb{P}_{n^*}$ the empirical distribution of bootstrapped predicted ILR curves. The lower and upper bounds of CLR can be obtained similarly from $S = \{\hat{c}^{1}_{n^*}(x),\ldots,\hat{c}^{B}_{n^*}(x)\}, x \in \xl $. For $s = 9$, since the forecast is one additional point, the depth-based methods essentially yield the point-wise prediction interval.  
    We consider extremal depth (EXD) \citep{extremaldepth} for $D(y, \cdot)$. 

Our method can be viewed as a semi-parametric hierarchical model:  
\begin{align}
y_{n}(x) \mid \beta_{n} &\sim \mathcal{N}\Big(\mu(x)+\sum_{k}\phi_{k}(x)\beta_{k,n}, 
\sigma^{2}(x)\Big), \\
\beta_{k,n}\mid X_{n} &\sim f_{k}(X_{n})+\xi_{k,n}.
\end{align}
Thus, uncertainty arises from both regression residuals and functional residuals. 
The bootstrap mimics this hierarchical structure by resampling both components.

\subsection{Fixed-Origin Backtesting}\label{sec:fixedAY}

To test out the forecast and bootstrap algorithm, we implement a fixed–origin backtesting procedure. 
Specifically, we fix the calendar year $T$ and treat accident year $T$ as ``current,'' 
with ILR curves available only up to lag $s-1$. 
We then forecast the continuation of accident year $T$ using our proposed method 
and compare against realized values once the full triangle is known. 
This exercise mimics the practical reserving task of predicting partially developed cohorts. We implement this workflow on three representative companies P52, C260 and P1406. Table \ref{tbl:covariates} lists their covariates.     

\begin{table}[!ht]
		\centering
		{\footnotesize 	\begin{tabular}{rrrrr}
				\toprule
				Company & Business Focus & Ownership Structure & Geographic Focus & Net Premium Earned in 2010 \\
				\midrule
				P53 & WkComp & Stock & West & 321,786,000 \\
				C260 & Personal & Mutual & South & 177,078,000 \\
                P1406 & Personal & Mutual & National & 310,728,000 \\
				\bottomrule
		\end{tabular}}
		\caption{Company covariates for P53, C260 and P1406, cf. Figure \ref{fig:ilrclr2010}. }
		\label{tbl:covariates}
	\end{table}

\paragraph{Setup.}
Let $\mathcal{D}_{\mathrm{train}}(s) = \{y_{n}(x): n \in \mathcal{N}, t \leq T-9, x \in \mathcal{X}\}$ 
denote the training set comprising fully developed curves up to accident year $T-9$. 
For the focal accident year $T$, we observe only $y_{i,T}(x)$ for $x \in \{0,\dots,s-1\}$. 
Denote the predicted continuation as $\hat{y}_{i,T}(x), x \in \{s,\dots,9\}$, 
with corresponding cumulative forecast $\hat{c}_{i,T}(9)$.

\paragraph{Sequential updating.}
We conduct the procedure for $s=1,\dots,9$. 
At each $s$, the following steps are performed:
\begin{enumerate}
    \item Estimate FPCA basis $\{\hat{\mu},\hat{\phi}_{k}\}$ and regression models $f_{k}$ 
    from $\mathcal{D}_{\mathrm{train}}(s)$.
    \item For accident year $T$, obtain regression prior scores 
    $\hat{\beta}^{\mathrm{RM}}_{k,i,T}$.
    \item Apply penalized least squares (\ref{eq:pls_matrix}) using the observed partial curve. 
    \item Compute forecasts $\hat{y}_{i,T}(x)$ and $\hat{c}_{i,T}(9)$.
\end{enumerate}
As $s$ increases, the algorithm is updated with longer partial curves and larger training sets, 
so forecast precision should improve monotonically.

\begin{figure}[!htb]
		\centering
            \begin{tabular}{ccc}
			\hspace*{-20pt}\includegraphics[width=0.325\textwidth]{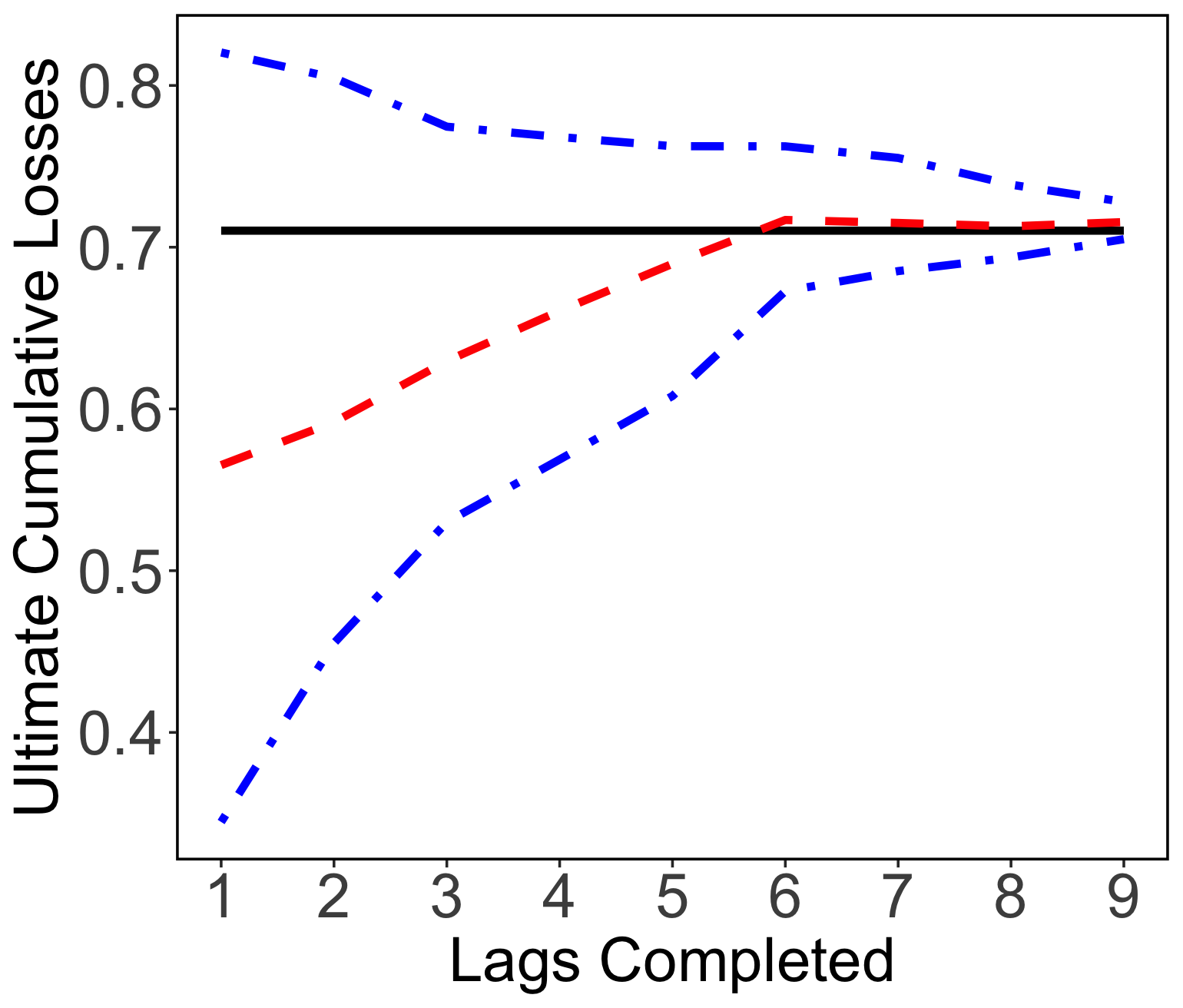} &
			\includegraphics[width=0.325\textwidth]{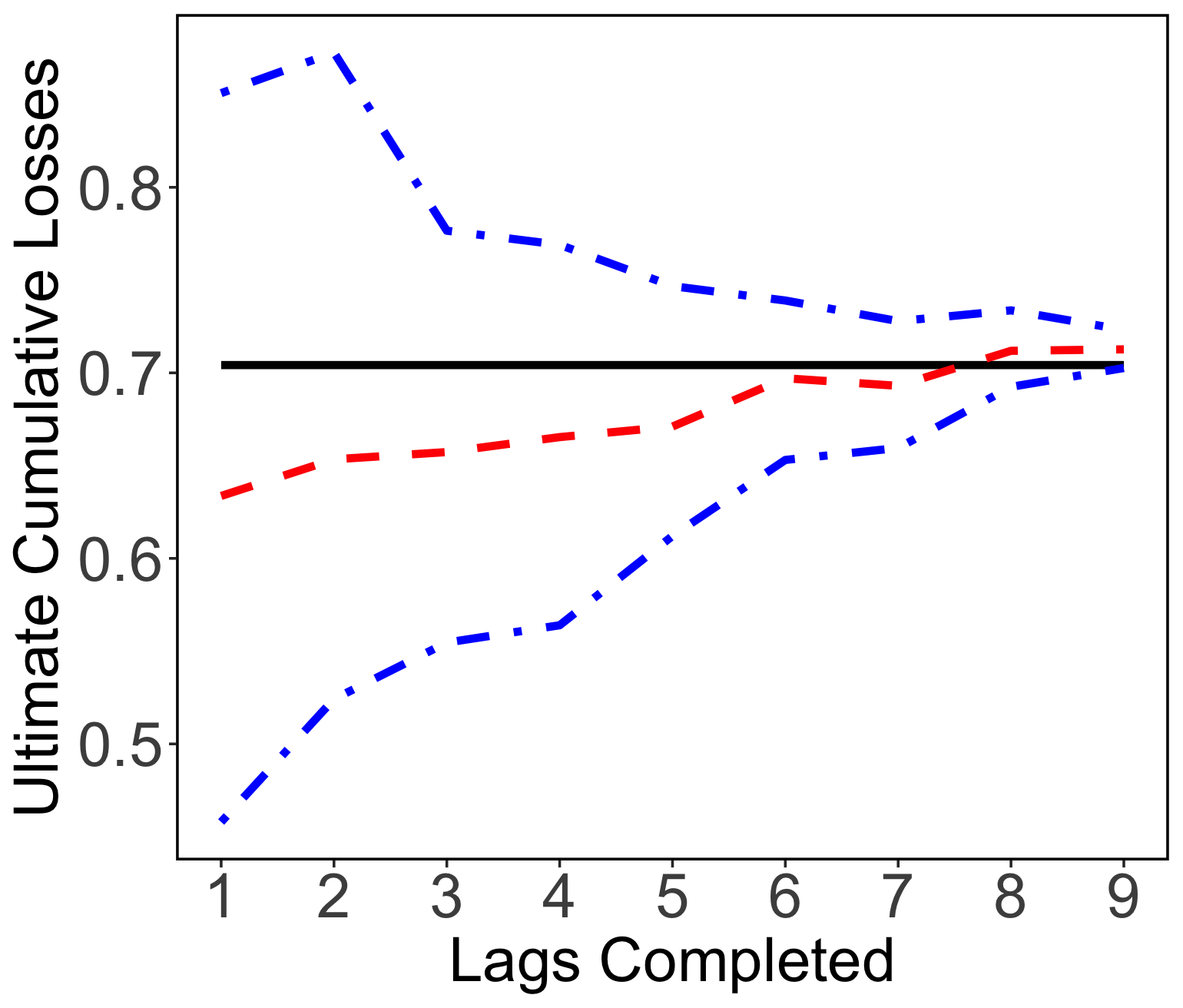} &
			\includegraphics[width=0.325\textwidth]{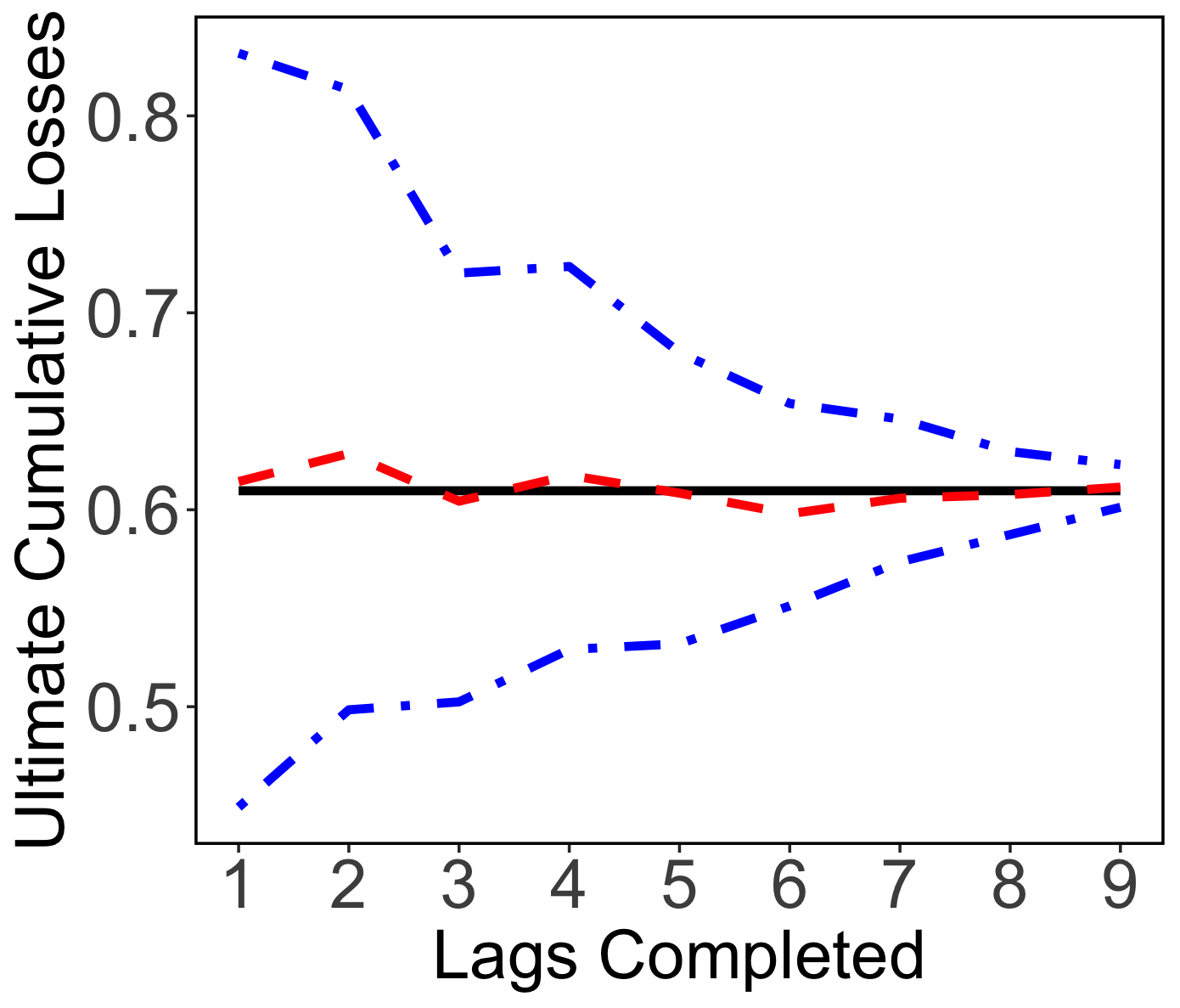} \\
            P53 & C260 & P1406 \\
            \end{tabular}
		\caption{Evolution of ultimate CLR forecast for AY 2010 as a function of developed lag $s$ for three representative triangles listed in Table \ref{tbl:covariates}. Black line--- ground truth ultimate CLR; red line ---mean forecast ultimate CLR; blue lines---95\% EXD prediction interval.}
        \label{fig:uclr}
	\end{figure}

\paragraph{Empirical findings.}
Figure~\ref{fig:uclr} illustrates typical results for three representative companies. 
As expected, forecasts are least accurate for $s=1$--$2$, where limited early development 
provides weak signal and predictions are dominated by regression priors. 
Accuracy and calibration improve steadily as $s$ increases, 
with predictive intervals narrowing around the realized CLR once $s \geq 5$. 
This confirms that our penalized least squares adjustment balances prior information 
with observed development, leading to stable and progressively accurate forecasts.

For the full evolution of the partial curve and future forecasts across development years for the three companies both in terms of ILR and CLR, see Figure~\ref{fig:ilrclr2010} in Appendix~\ref{app:plots}.

\subsection{Application to Loss Triangle Completion}\label{sec:illustrate}

We collect loss history from $I = 224$ companies from the earliest accident year 1987 labeled as $t = 1$ to the most recent accident year 2010 labeled as $T = 24$. We train the PLS model on $y_{i,t}(x), t \le T-9$ and features $X_{i,t}$ including accident year $t$, earned premium $P_{i,t}$ and time-invariant company and business line characteristics. We then forecast the lower right triangle $y_{i,t}(x), x > T - t$ using the model and upper left triangle $y_{i,t}(x), 0 \le x \le T - t$ and evaluate the performance against the held-out true value of the lower right triangle. Before carrying out the PCA and regression, we remove the outliers, namely curves with outlying bandwidth depth (BD) scores. A total of 50 outlying ILR curves are removed, see Figure \ref{fig:fdaoutlier}.  

The procedure is similar to the previous section, but we start from accident year 2002, or $t = 16$, with $s = 9$, forecast the remainder of the ILR curves, augment the training set with the completed ILR curves, and update $s$ to $s-1$. We then obtain the CLR and multiply by earned premium to recover prediction for cumulative losses for comparison with traditional reserving methods.

We report the optimal $K(s)$ and $\lambda(s)$ tuning results in Table \ref{tbl:K-lam4}. The collection of loading vectors used in the model can be found in Table \ref{tab:pca_loadings} in Appendix~\ref{app:tables}. The coefficients of LASSO regression can be found in Table \ref{tab:lasso_full} in Appendix~\ref{app:tables}. We observe that the LASSO uses all the covariates for shorter lags $s < 7$ and drops most covariates for $s=7,8,9$.

\begin{table}[!htb]
        \centering
        \begin{tabular}{ccccc}
            \toprule
            Accident Year (Lag $s$) & Training Curves & Optimal $K(s)$ & Optimal $\lambda(s)$ & MAPE \\
            \midrule
            2002 \ \ (9) & 1550 & 8 & 0.0471 & 0.0065 \\
            2003 \ \  (8) & 1710 & 8 & 0.0853 & 0.0117 \\
            2004 \ \ (7) & 1882 & 6 & 0.0387 & 0.0192 \\
            2005 \ \ (6) & 2063 & 6 & 0.1111 & 0.0272 \\
            2006 \ \ (5) & 2251 & 5 & 0.0898 & 0.0396 \\
            2007 \ \ (4) & 2445 & 4 & 0.1000 & 0.0557 \\
            2008 \ \ (3) & 2653 & 3 & 0.1606 & 0.0820 \\
            2009 \ \ (2) & 2866 & 2 & 0.1762 & 0.1284 \\
            2010 \ \ (1) & 3088 & 1 & 0.2271 & 0.1825 \\
            \bottomrule
        \end{tabular}
        \caption{Optimal number of PCA factors $K(s)$ and penalty $\lambda(s)$ as a function of number of completed lags $s$.}
        \label{tbl:K-lam4} 
\end{table}

\begin{figure}[!ht]
    	\centering
        \begin{subfigure}[b]{0.32\textwidth}
    		\includegraphics[width=\linewidth]{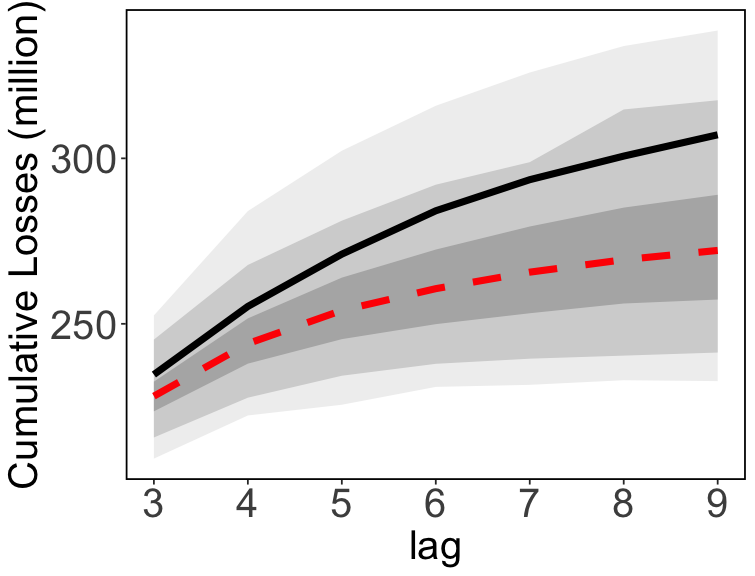}
    		\caption{P53: $s=3$}
    		\label{p53s3}
    	\end{subfigure} 
	    \begin{subfigure}[b]{0.32\textwidth}
    		\includegraphics[width=\linewidth]{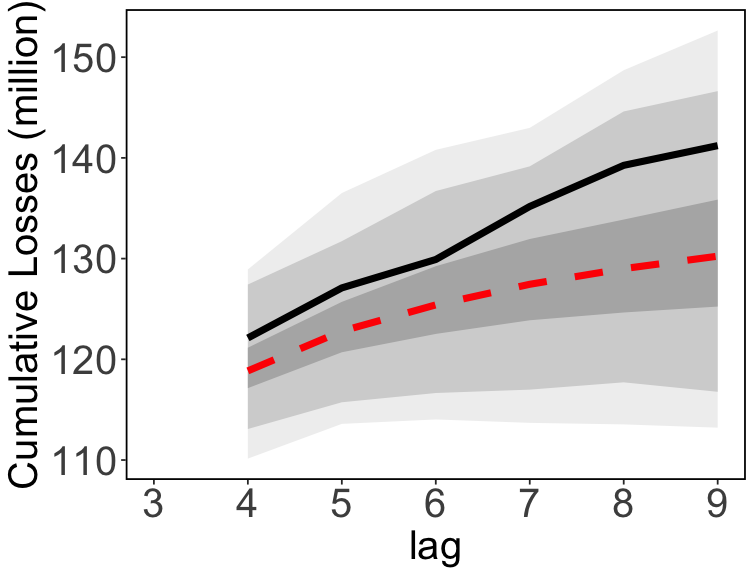}
    		\caption{C260: $s=4$}
    		\label{p53s3}
    	\end{subfigure}
        \begin{subfigure}[b]{0.32\textwidth}
    		\includegraphics[width=\linewidth]{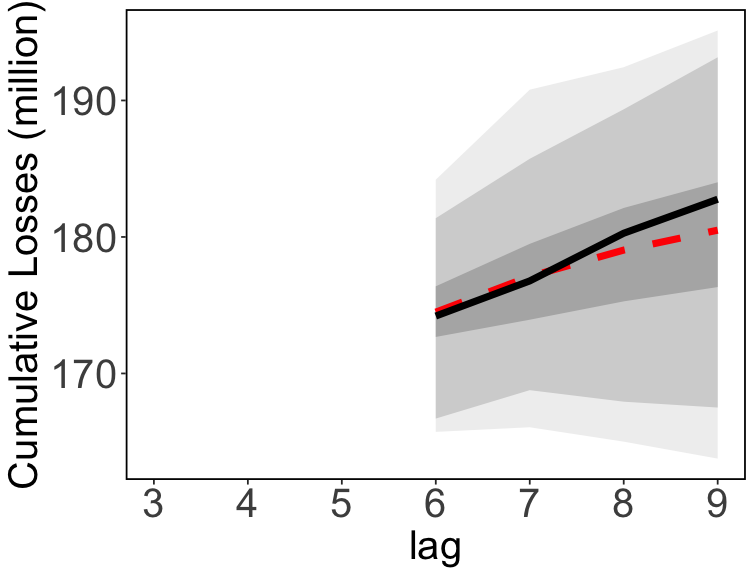}
    		\caption{P1406: $s=6$}
    		\label{p1406s6}
    	\end{subfigure}
	\caption{Illustrating PLS predictions for cumulative losses and respective EXD-based 50\%, 90\% and 95\% functional regions (grey shades). Solid black lines: realized cumulative losses (million USD); dashed red lines: predicted cumulative losses.}
	\label{pls-pred-bd}
    \end{figure}

In Figure \ref{pls-pred-bd} we show the resulting forecasts and bootstrapped intervals for several confidence levels. In the left and center panel, the predicted CLR is below the true CLR, but both are covered in the 90\% EXD central region. The right panel where more lags are complete, shows more accurate prediction and the true CLR lies in the 50\% EXD central region. In the next section, we will formally make assessment on the accuracy and precision of the forecast.
	
\section{Forecast Assessment}\label{sec:assessment}

Having introduced the methodology, we now assess its empirical performance in forecasting 
loss development. Evaluation is performed along three dimensions: (i) point forecast accuracy, 
(ii) calibration and sharpness of probabilistic forecasts, and (iii) functional coverage of 
CLR trajectories. We benchmark our approach against the classical 
Mack chain-ladder (CL) model (\citealp{m93}; see also \citealp{wm08}), 
the most widely used reserving method in practice.  

\subsection{Experimental Design}
We consider a subset of 137 insurers with fully observed triangles from accident years 
2001--2010. This ensures availability of ``future truth'' for validating predictions. 
For each accident year $t\in\{T-8,\dots,T\}$, and for each development horizon $s=1,\dots,9$, 
we fit the models using only information up to lag $s-1$ and then forecast the continuation 
of ILR and CLR curves.  Our evaluation metrics include:
\begin{enumerate}
    \item \textbf{MAPE of ultimate cumulative losses}
    \[
    \mathrm{MAPE}_t = \frac{1}{I}\sum_{i=1}^{I}
    \left|\frac{\hat{c}_{i,t}(9)-c_{i,t}(9)}{c_{i,t}(9)}\right|, \;\; t \in [T-8, T]
    \]
    \item \textbf{Empirical coverage of prediction intervals}, i.e., the fraction of triangles where the realized value falls within the $(1-\alpha)$ forecast band.
    \[
    \mathrm{Coverage}_{\alpha,t} = \frac{1}{I}\sum_{i=1}^{I} 
    \mathbf{1}\{L_{i,t} \leq c_{i,t}(9) \leq U_{i,t}\}, \;\; t \in [T-8, T]
    \]
    where $[L_{i,t},U_{i,t}]$ is the $(1-\alpha)$ predictive interval for ultimate CLR. 
    \item \textbf{Functional coverage} of CLR curves, assessed via depth-based envelopes (\citealp{extremaldepth, dai2020}).  
    \[
    \mathrm{FuncCoverage}_{\alpha,t}(s) = \frac{1}{I}\sum_{i=1}^{I} 
    \mathbf{1}\{c_{i,t}(x) \in C_{1-\alpha}\}, \; x \in X_l, t \in [T-8, T]
    \]
    $C_{1-\alpha}$ is the depth-based prediction interval as in Equation \eqref{eq:depth-region}.
    \item \textbf{Interval Score} (see \cite{GneitingRaftery2007}), which jointly evaluates 
    coverage and sharpness. A general formula of interval score of true value $Y$ and prediction upper bound $U$ and lower bound $L$ is
    \[
    IS_{\alpha}(L,U;Y) = (U-L) + \tfrac{2}{\alpha}(L-Y)\mathbf{1}\{Y<L\}
    + \tfrac{2}{\alpha}(Y-U)\mathbf{1}\{Y>U\}.
    \]
    We will evaluate the interval score for ultimate cumulative losses and the predicted cumulative loss curve.
    
\end{enumerate}
This combination of metrics follows best practice in forecast evaluation 
(\citealp{Dawid1984, DieboldMariano1995, Patton2011}). These metrics jointly assess accuracy and calibration, and allow comparison against 
benchmarks such as Mack’s chain-ladder (\citealp{m93}), 
Gaussian process reserving models (\citealp{ludkovski2020gaussian, ang2022hierarchical}), 
and machine learning approaches (\citealp{wuthrich18, schneider2025advancing}).

\subsection{Point Forecast Accuracy}
Table~\ref{tbl:mape-coverage} reports MAPE for ultimate cumulative losses. 
We find that CL tends to perform better for short development horizons ($s=1$--$4$), 
where the information content of the partial curve is limited and CL’s multiplicative 
structure is effective. However, our functional PLS method outperforms CL once more lags 
are observed ($s \geq 5$), reflecting its ability to exploit the functional similarity across 
triangles and covariate information. 

\begin{table}[!htb]
    \centering
    \begin{tabular}{c|rr|ccc|ccc}
        \toprule
        Lag 
        & \multicolumn{2}{c|}{MAPE} 
        & \multicolumn{3}{c|}{Coverage \%} 
        & \multicolumn{3}{c}{Interval score $\overline{UIS}_{0.95}$} \\
        $s$ & PLS & CL & PLS & EXD & CL & PLS & EXD & CL \\
        \midrule
        1 & 21.86\% & 16.53\% & 94.16\% & 89.78\% & 86.13\% & 0.7132 & 1.0490 & 0.7098 \\
        2 & 13.11\% & 10.31\% & 98.54\% & 97.08\% & 79.56\% & 0.4857 & 0.4286 & 1.0422 \\
        3 & 9.06\%  & 7.21\%  & 98.54\% & 93.43\% & 71.53\% & 0.2969 & 0.5009 & 0.9904 \\
        4 & 5.31\%  & 5.18\%  & 98.54\% & 96.35\% & 70.80\% & 0.2379 & 0.1854 & 0.4047 \\
        5 & 3.90\%  & 4.10\%  & 99.27\% & 99.27\% & 73.72\% & 0.1772 & 0.1434 & 0.1808 \\
        6 & 2.38\%  & 2.69\%  & 100.00\% & 98.54\% & 79.56\% & 0.1254 & 0.0894 & 0.1176 \\
        7 & 1.80\%  & 2.07\%  & 97.81\% & 97.81\% & 84.67\% & 0.1263 & 0.1208 & 0.0900 \\
        8 & 1.14\%  & 1.41\%  & 97.81\% & 97.81\% & 83.94\% & 0.0586 & 0.0561 & 0.0669 \\
        9 & 0.56\%  & 0.67\%  & 94.89\% & 94.89\% & 92.70\% & 0.0299 & 0.0299 & 0.0354 \\
        \bottomrule
    \end{tabular}
    \caption{Statistical scoring metrics for forecasting ultimate cumulative losses. We report MAPE, coverage at 95\% level, and average ultimate interval scores $\overline{UIS}_{\alpha}$ with $\alpha=0.95$. PLS refers to point-wise interval, CL refers to chain ladder, and EXD to extremal depth.}
    \label{tbl:mape-coverage}
\end{table}

The average interval scores of ultimate cumulative losses based on $s$ developed lags are calculated by weighing each curve by the size of the business line (earned premium),
\begin{align}
  \overline{UIS}_\alpha(s) :=  \frac{\sum_i IS_\alpha(C_{i,t}(9)) P_{i,t}}{\sum_i P_{i,t}}.
\end{align}
The two right-most columns of Table \ref{tbl:mape-coverage} compare the PLS method with point-wise and EXD-based intervals against the \citet{m93} chain ladder method in terms of the ultimate interval score $\overline{UIS}_{0.95}$ on 137 test triangles. The point-wise or EXD method has a favorable interval score for all lags except for lags 1 and 7.

The results reveal a clear trade-off: CL yields slightly lower MAPE at early lags, 
but significantly underestimates predictive uncertainty, as seen in low coverage rates. 
In contrast, our functional method provides consistently higher coverage 
and lower interval scores, indicating better-calibrated probabilistic forecasts.

\subsection{Calibration and Predictive Distribution}
\begin{figure}[H]
	 	\centering
	 	\includegraphics[width=0.98\linewidth]{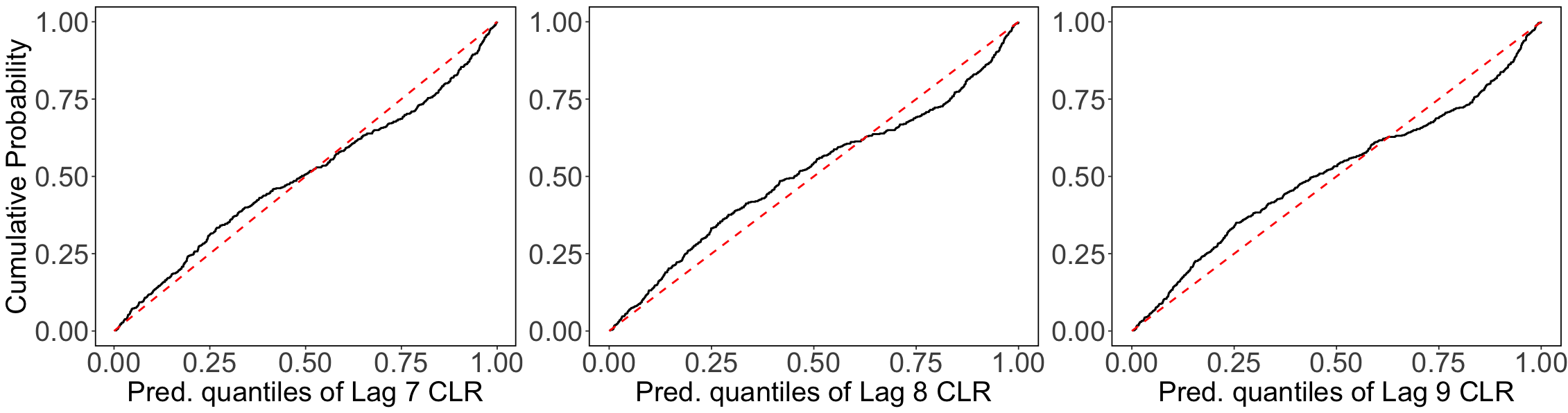}
	 	\caption{Empirical CDF comparing realized cumulative loss ratios to predicted quantiles across the 137 triangles for 3-6 lags complete.}
 		\label{fig:ecdf}
 	\end{figure}

Figure~\ref{fig:ecdf} shows the empirical distribution of realized CLR 
quantiles relative to our forecast distributions (probability integral transforms). 
If forecasts are well calibrated, this empirical CDF should be close to uniform 
\citep{Dawid1984,wilks1990combination,denuit2021autocalibration}. We observe that the functional method produces distributions of quantiles that are
close to uniform, while CL exhibits systematic undercoverage, 
especially for intermediate lags. This diagnostic confirms that CL’s 
variance assumptions underestimate true uncertainty, 
while our bootstrap-based functional forecasts are more reliable.  

Figure \ref{fig:ecdf} shows the combined lag 7-9 CLR predictive results for accident years 2005-2008 (with 6,5,4,3 lags complete respectively). The reverse S-shape in all three panels suggests the predictions tend to be at either lower or higher quantiles of the distribution. Further test of uniformity of the realized quantiles is done by the Kolmogorov Smirnov (KS) statistic. The KS score for the three ECDF is $0.0741, 0.0869$, and $0.0954$. Based on the critical value $1.35/\sqrt{137 \cdot 4} = 0.0577$, we reject the null hypothesis and the predictive distribution quantile is not uniform.

\subsection{Functional Coverage of CLR Curves}
Beyond pointwise metrics, we evaluate the coverage of entire CLR trajectories. 
We compute functional prediction envelopes using extremal depth (EXD), 
which provide simultaneous bands over development lags (\citealp{extremaldepth}). 
Table~\ref{tbl:functional-score} reports functional coverage rates and 
cumulative interval scores (CIS), defined as
\begin{equation}
CIS_{\alpha}(s) = \frac{1}{I}\sum_{i=1}^{I} \frac{1}{10-s}
\sum_{j=s}^{9} IS_{\alpha}\big(L_{i,t}^{(s)}(j),U_{i,t}^{(s)}(j);c_{i,t}(j)\big).
\end{equation}
We find that CL consistently delivers coverage well below the nominal 95\% level, 
especially for $s<6$, whereas our method achieves coverage close to nominal, 
with tighter scores for $s\geq 4$.  

\begin{table}[!htb]
    \centering
    \begin{tabular}{c|ccc|ccc}
        \toprule
        \multirow{2}{*}{$s$} 
        & \multicolumn{3}{c|}{Coverage (\%) at $\alpha=95\%$} 
        & \multicolumn{3}{c}{Interval Score } $\overline{CIS}_{0.95}$ \\
        & PLS & EXD & CL & PLS & EXD & CL \\
        \midrule
        1 & 70.80\% & 72.26\% & 82.48\% & 0.4813 & 0.6031 & 0.5125 \\
        2 & 96.35\% & 95.62\% & 72.26\% & 0.3465 & 0.3241 & 0.4802 \\
        3 & 91.24\% & 89.78\% & 64.23\% & 0.2005 & 0.2177 & 0.4728 \\
        4 & 94.89\% & 94.16\% & 66.42\% & 0.1589 & 0.1331 & 0.2003 \\
        5 & 95.62\% & 97.08\% & 70.07\% & 0.1188 & 0.1092 & 0.1121 \\
        6 & 96.35\% & 95.62\% & 75.91\% & 0.0906 & 0.0769 & 0.0734 \\
        7 & 97.81\% & 97.81\% & 78.10\% & 0.0870 & 0.0877 & 0.0573 \\
        8 & 95.62\% & 97.08\% & 82.48\% & 0.0476 & 0.0474 & 0.0501 \\
        9 & 94.89\% & 94.89\% & 92.70\% & 0.0299 & 0.0299 & 0.0354 \\
        \bottomrule
    \end{tabular}
    \caption{Statistical scores for forecasting cumulative loss curves from given lag $s$ to lag 9. We report coverage at 95\% level, and average cumulative interval scores $\overline{CIS}_{\alpha}$ with $\alpha=0.95$. PLS refers to point-wise interval, CL to chain ladder and EXD to extremal depth.}
    \label{tbl:functional-score}
\end{table}

Taken together, these results highlight three conclusions. 
First, CL achieves good point accuracy in early lags, but its variance assumptions 
lead to systematically under-dispersed predictive distributions. 
Second, our functional method achieves competitive or superior accuracy 
at later lags, while providing better-calibrated uncertainty quantification. 
Third, functional coverage metrics show that entire CLR trajectories are captured 
with higher reliability by our approach.  
From a practical perspective, this implies that regulators and insurers can use 
our forecasts not only for reserve point estimates but also for realistic predictive envelopes, 
supporting solvency analysis and capital adequacy under uncertainty.

\begin{figure}[!ht]
	\centering
	\includegraphics[width=0.32\linewidth]{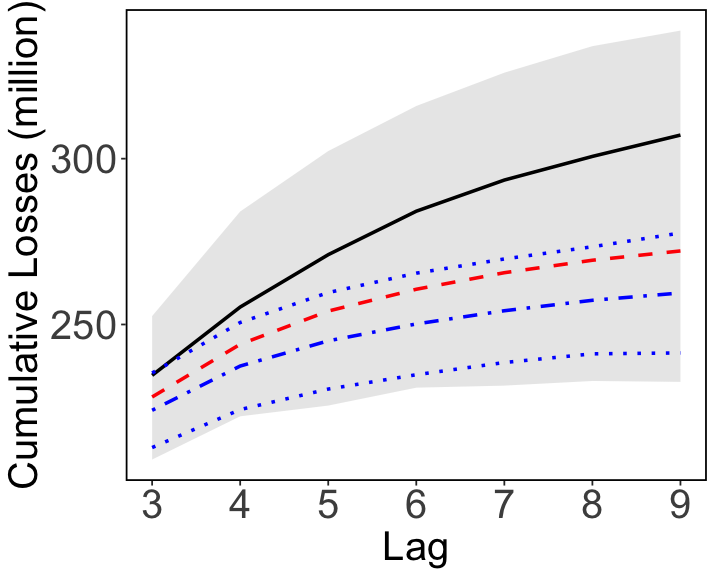}
    \includegraphics[width=0.32\linewidth]{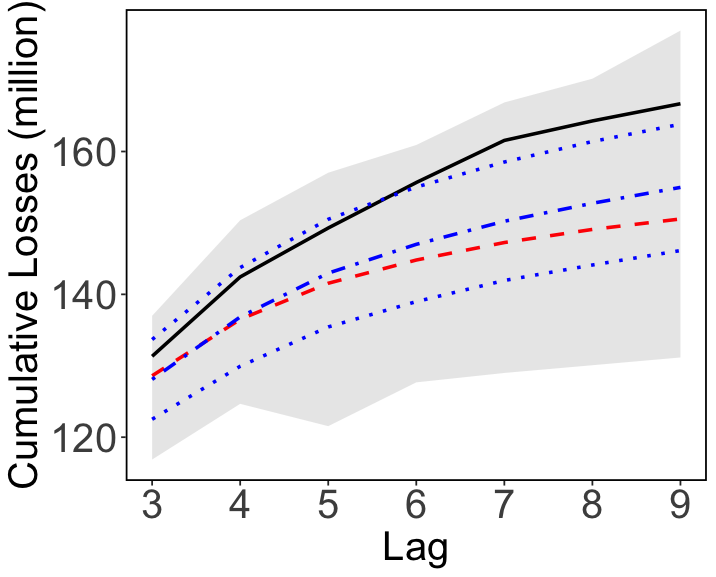}
    \includegraphics[width=0.32\linewidth]{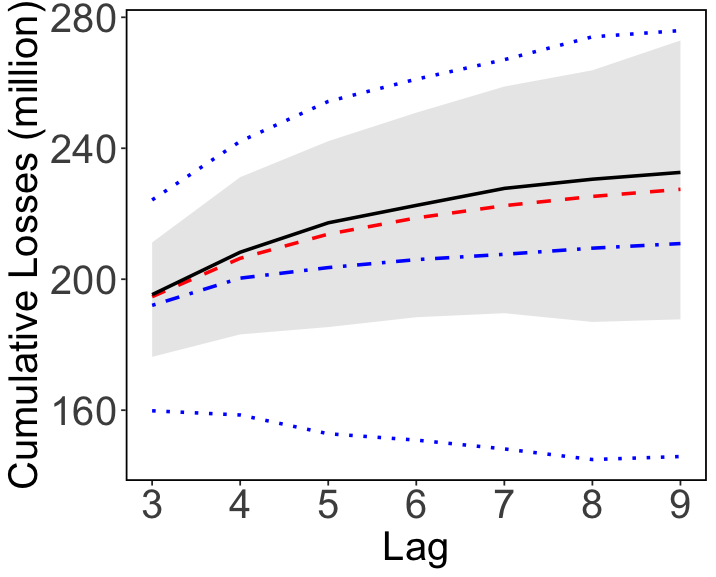} \\[0.5em]
    \includegraphics[width=0.32\linewidth]{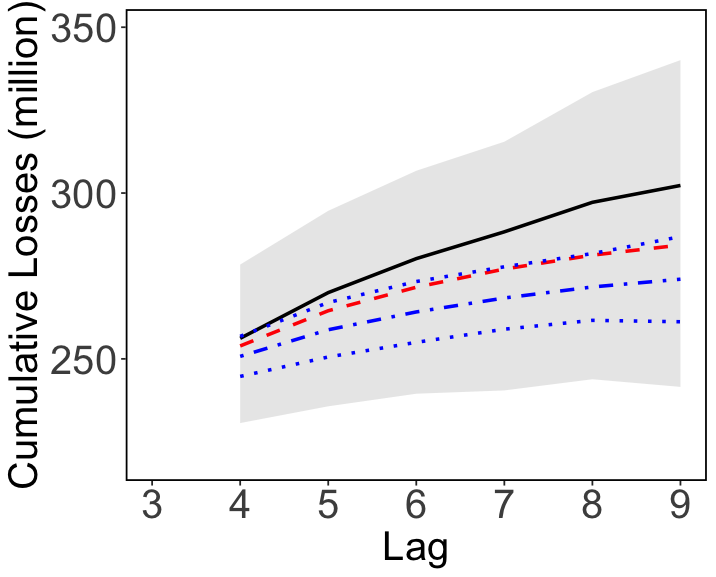}
    \includegraphics[width=0.32\linewidth]{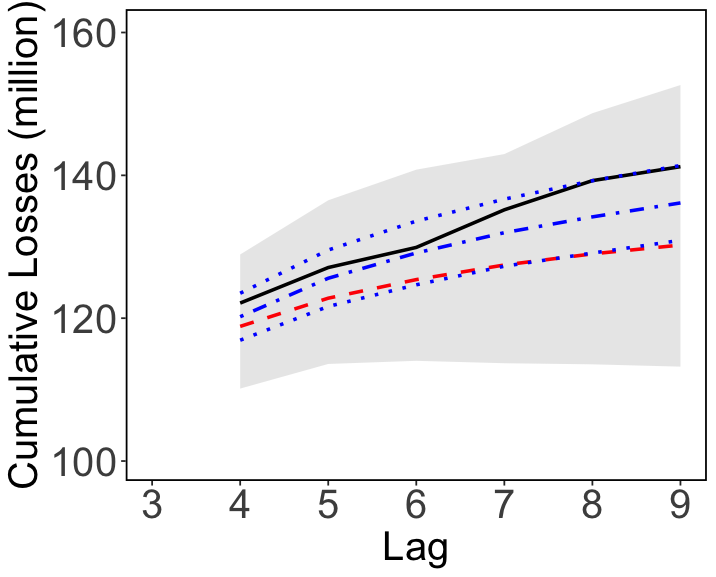}
    \includegraphics[width=0.32\linewidth]{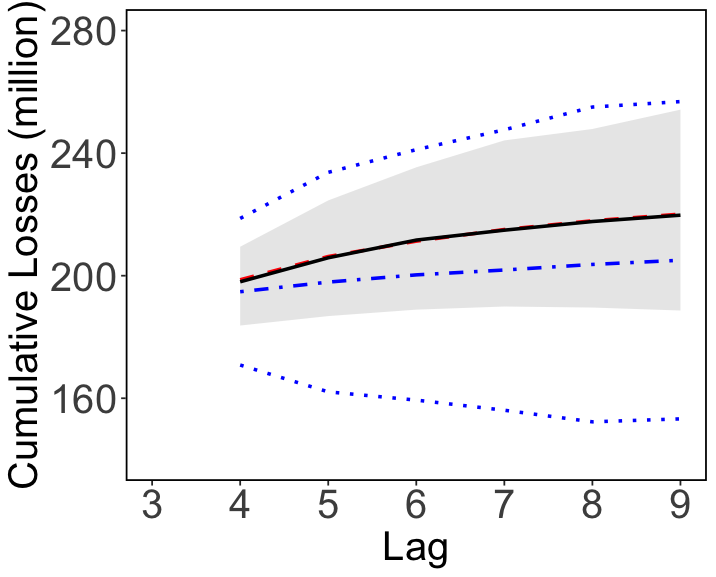} \\[0.5em]
    \includegraphics[width=0.32\linewidth]{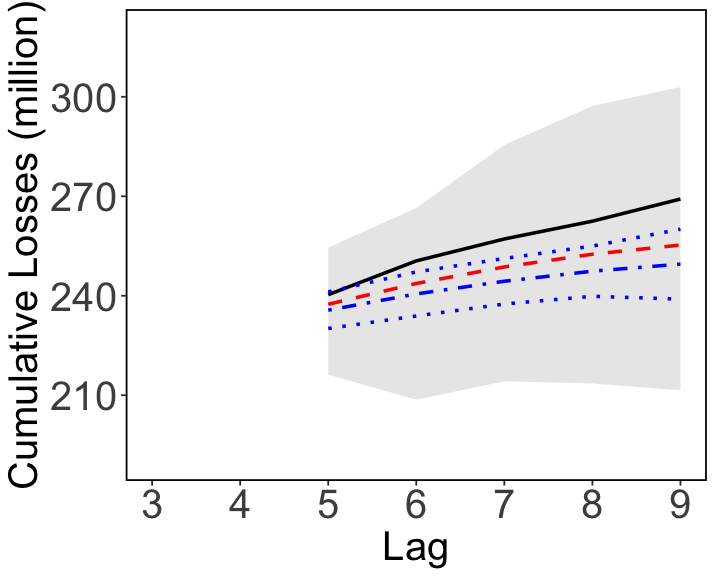}
    \includegraphics[width=0.32\linewidth]{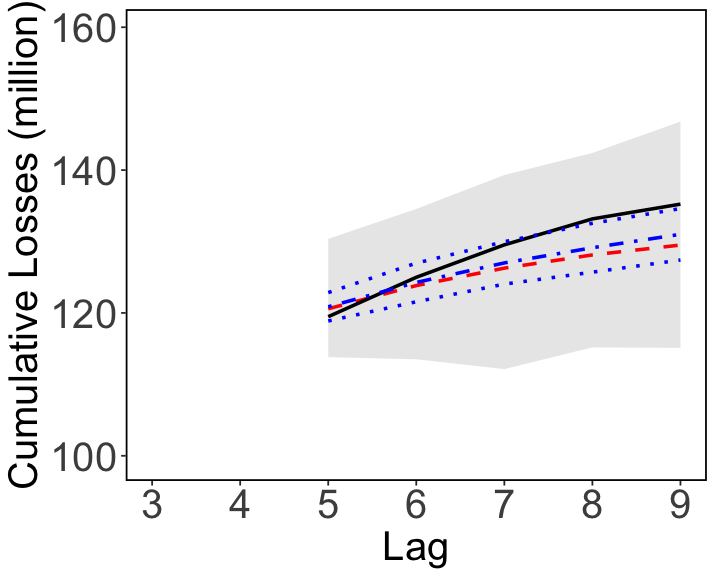}
    \includegraphics[width=0.32\linewidth]{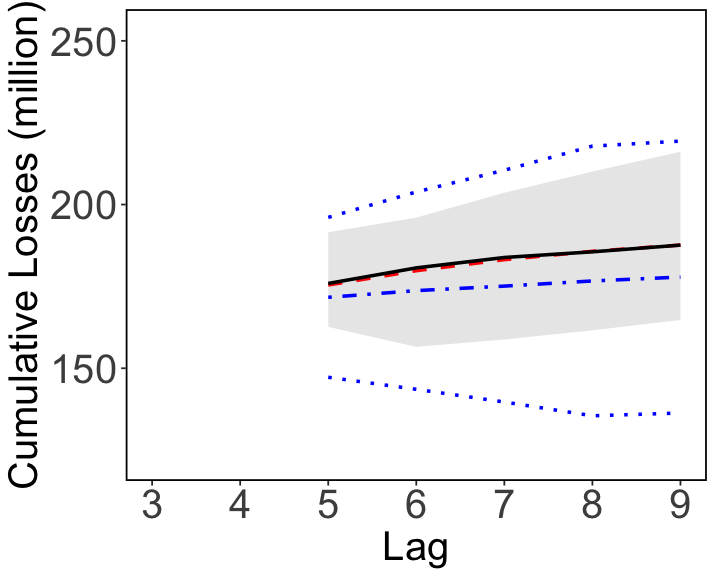} \\[0.5em]
    \begin{subfigure}[b]{0.32\textwidth}
    	\includegraphics[width=\linewidth]{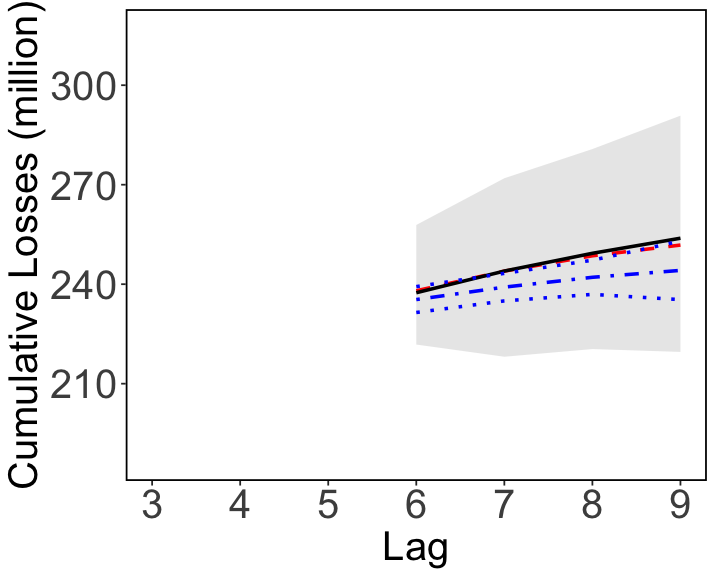}
    	\caption{P53}
    	\label{p53}
    \end{subfigure} 
	\begin{subfigure}[b]{0.32\textwidth}
    	\includegraphics[width=\linewidth]{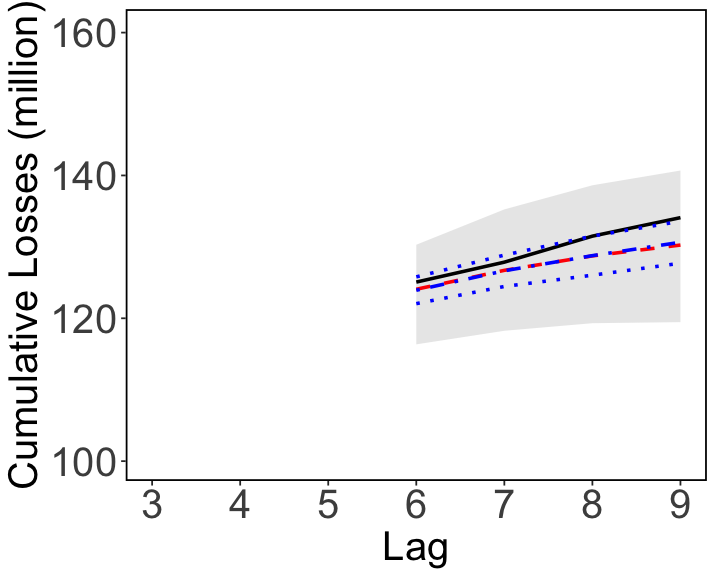}
    	\caption{C260}
    	\label{c260}
    \end{subfigure}
    \begin{subfigure}[b]{0.32\textwidth}
    	\includegraphics[width=\linewidth]{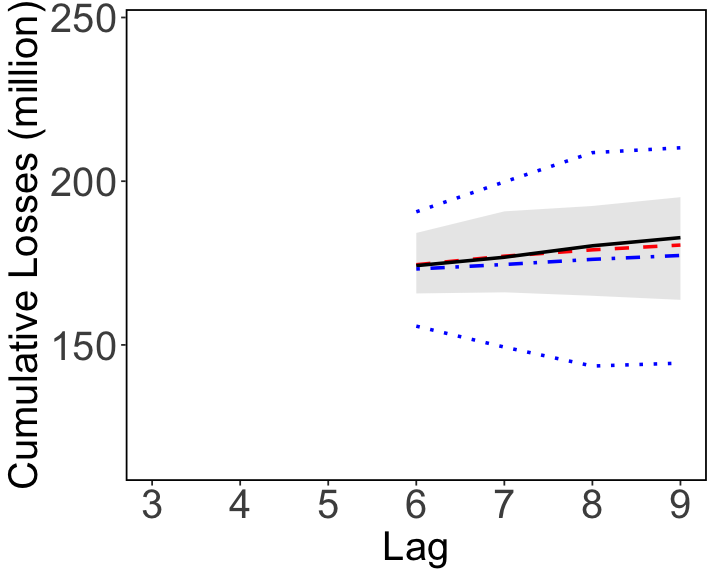}
    	\caption{P1406}
    	\label{p1406}
    \end{subfigure}
	\caption{Comparing PLS prediction and EXD coverage against chain-ladder method. Red curves: PLS prediction; black curves: realized actual ILRs; blue lines: chain ladder predictions and 95\% prediction intervals; grey intervals: 95\% EXD coverage region.}
    \end{figure}

    We next show a few illustrative predictions and functional envelopes of PLS versus CL. Figure \ref{p53} shows the case where the PLS method gives better prediction than chain ladder and sufficiently wide predictive intervals. Figure \ref{c260} show the case where PLS method gives worse prediction, but sufficiently covers the true curve, while CL does not cover the realized losses. Figure \ref{p1406} show the case where PLS method gives better prediction and tighter but sufficient prediction interval. We also note that removing outlier curves prior to running the model helps with improving the MAPE and interval score, cf.~Table \ref{tbl:mape-coverage-withoutlier} and \ref{tbl:functional-score-withoutlier} in Appendix~\ref{app:tables}.

\subsection{Extensions to Other Domains}
\label{sec:mortality}

While our empirical study focused on P\&C loss development, the framework is general. For example, epidemiology modeling, or mortality analysis, also involves incomplete cohort structures.

\subsubsection*{Epidemiology}

Beyond insurance applications, the proposed functional–probabilistic framework naturally extends to epidemiology, where delayed reporting structures produce data highly analogous to run-off triangles. In infectious disease surveillance, cases are often registered with a lag between infection, diagnosis, and reporting, generating incomplete ``reporting triangles'' \citep{lawless1994adjustments,hohle2014bayesian}. This setting raises similar methodological challenges as loss development: incomplete curves, strong dependence across development lags, and the presence of reporting anomalies. A concrete illustration is provided by the analysis of acquired immune deficiency syndrome (AIDS) in England and Wales, where reported case counts also form incomplete development triangles with a fixed 12-quarter horizon, structurally identical to the loss triangles considered here \citep{Angelis2018}.


Let $t \in \mathcal{T}$ denote calendar time (e.g., diagnosis week) and $x \in \mathcal{X}$ the reporting delay (lag). Define $C_{t}(x)$ as the cumulative number of cases diagnosed in period $t$ that are reported within $x$ weeks. The incremental counts are
\[
Y_{t}(x) := C_{t}(x)-C_{t}(x-1), \quad x \geq 1, 
\]
with $Y_{t}(0) = C_{t}(0)$ denoting the immediately reported cases. Analogous to incremental loss ratios, we may normalize by exposures or population size to obtain scale-free curves $y_{t}(x)$, suitable for comparison across regions or age groups.

The reporting triangle $\{Y_{t}(x): t \leq T, \, 0 \leq x \leq T-t\}$ is then structurally identical to insurance loss triangles. Forecasting unreported cases corresponds to predicting the lower-right corner of the reporting triangle. Traditional approaches include chain-ladder-type estimators \citep{hohle2017surveillance} or Bayesian hierarchical models \citep{stoner2020hierarchical}. Our functional data analysis (FDA) approach provides a nonparametric alternative: case-delay profiles are treated as functional curves, decomposed via FPCA, with partially observed trajectories completed through penalized least squares and bootstrap-based predictive envelopes.

Epidemiological reporting is further complicated by systemic anomalies, such as reporting delays during holidays or sudden diagnostic surges during epidemics. These correspond to outliers in our FDA framework. Depth-based detection methods \citep{sun2012exact,extremaldepth} can robustify the analysis by down-weighting atypical years (e.g., pandemic waves) when estimating mean reporting patterns. This mirrors our treatment of anomalous insurers in Section \ref{sec:outlier-detection}.

Forecasting incomplete cohorts of infected individuals can be viewed through a cohort--development lens. For a birth-cohort analogue, consider infection cohorts $c = t-x$; right-truncated infection cohorts (those diagnosed recently) resemble open triangles in reserving. Illness–death models formulated in Lexis space provide another powerful representation of such cohort dynamics \citep{brinks2014illness}, and age–period–cohort regression on the Lexis diagram is a natural extension \citep{carstensen2007age}. Estimation of delay distributions via robust functional regression allows prediction of unreported cases and quantification of associated uncertainty. Probabilistic envelopes provide epidemiologists with calibrated real-time case estimates, an important input for policy responses during outbreaks.

In summary, epidemiological reporting triangles share the same mathematical structure as insurance loss triangles. By adapting our functional–probabilistic methodology, we would obtain flexible, robust forecasts of unreported cases, enhancing both public health surveillance and real-time epidemic response \citep{bstos2019reporting}.

\subsubsection*{Mortality}

Another natural domain of application is cohort-driven mortality modeling. Age-specific death rates arranged on a Lexis diagram give rise to incomplete birth cohorts that mirror right-truncated loss triangles. The functional–probabilistic methods developed here can therefore be used to forecast partially observed age–cohort mortality profiles, in close analogy to how we complete insurance development curves. This parallel has been widely recognized in the actuarial literature \citep{GanLin2015IME,HuntVillegas2015IME,TsaiKim2022IME}, reinforcing the unity of reserving problems across nonlife insurance and demography.

\section{Conclusion}\label{sec:conclusion}

This paper introduced a functional perspective on the analysis and forecasting of loss development in property and casualty insurance. By reinterpreting run-off triangles as collections of incremental and cumulative loss ratio curves, we moved beyond the traditional cell-based reserving paradigm and established a framework that leverages modern functional data analysis (FDA). This reconceptualization provides three key advances: (i) robust detection of outliers and anomalies, (ii) systematic incorporation of insurer- and market-level covariates, and (iii) probabilistic forecasting of incomplete curves with calibrated uncertainty quantification.

Empirically, we documented heterogeneous loss development patterns across more than one hundred U.S. workers’ compensation insurers over 24 accident years. Functional data depth allowed us to identify central development shapes and to flag atypical curves arising from data errors, unusual claims, or structural portfolio shifts. We showed that outliers are concentrated in longer development horizons and that company size is a major driver of heterogeneity, with smaller insurers exhibiting greater volatility and larger insurers displaying longer-tailed runoff. We also highlighted cyclical calendar-year effects aligned with well-known underwriting cycles, underlining the importance of accounting for market-wide shocks when benchmarking development patterns.

On the forecasting side, our methodology combines functional principal component analysis with regression-based priors and penalized curve completion. This approach allows partial curves to be extrapolated in a principled manner, balancing historical cross-sectional similarity with company-specific covariates. Forecast evaluation against the industry standard Chain-Ladder model \citep{m93,wm08} demonstrated that while CL performs competitively at very short horizons, our functional method dominates once a modest number of development lags are observed. In particular, it delivers sharper yet well-calibrated predictive intervals, as confirmed by interval scores \citep{GneitingRaftery2007} and functional coverage diagnostics. This improvement reflects the dual strength of FDA: its ability to exploit the global shape of development curves and its flexibility to integrate heterogeneous information sources.

Beyond actuarial reserving, the functional–probabilistic framework generalizes to other domains characterized by incomplete cohort data. We illustrated parallels with mortality forecasting 
where Lexis diagrams yield truncated age–cohort structures analogous to open triangles and epidemiological surveillance data organized into reporting triangles, with development corresponding to delayed case registration. r
These connections emphasize that the actuarial problem of projecting evolution of incurred-but-not-reported claims is part of a broader class of statistical forecasting problems on partially observed functional data.

Looking forward, several extensions are promising. First, the regression step can be enriched with additional financial or operational covariates, including external macroeconomic indicators, to capture drivers of long-tailed liabilities. Second, alternative resampling and conformal prediction methods \citep{graziadei2023conformal,hong2025conformal} can further strengthen distributional calibration. Third, linking the FDA approach with machine learning architectures \citep{kuo2019,schneider2025advancing} opens a path toward hybrid models that balance interpretability with predictive power. Finally, broader comparative studies across multiple lines of business and international datasets would help validate the robustness and generality of our framework.

In conclusion, reframing loss development as a functional forecasting problem yields both theoretical and practical gains. Our empirical study demonstrates that functional methods not only improve predictive performance relative to the Chain-Ladder benchmark but also provide a unified language for detecting anomalies, quantifying uncertainty, and extending actuarial techniques to adjacent fields. We hope that this work stimulates further cross-pollination between actuarial science, epidemiology, and demography, and contributes to the development of forecasting methodologies that are both statistically rigorous and operationally relevant.

\bibliography{document}

\begin{thebibliography}{56}
\newcommand{\enquote}[1]{``#1''}
\providecommand{\natexlab}[1]{#1}
\providecommand{\url}[1]{\texttt{#1}}
\providecommand{\urlprefix}{URL }
\providecommand{\bibAnnoteFile}[1]{%
  \IfFileExists{#1}{\begin{quotation}\noindent\textsc{Key:} #1\\
  \textsc{Annotation:}\ \input{#1}\end{quotation}}{}}
\providecommand{\bibAnnote}[2]{%
  \begin{quotation}\noindent\textsc{Key:} #1\\
  \textsc{Annotation:}\ #2\end{quotation}}

\bibitem[{Abdallah and Wang(2023)}]{abdallah2023rank}
Abdallah, Anas and Lan Wang (2023), \enquote{Rank-based multivariate {S}armanov for modeling dependence between loss reserves.} \emph{Risks}, 11, 187.
\bibAnnoteFile{abdallah2023rank}

\bibitem[{Ang and Lee(2022)}]{ang2022hierarchical}
Ang, Zi~Qing and See~Keong Lee (2022), \enquote{Hierarchical {B}ayesian {G}aussian process regression model for loss reserving using combinations of squared exponential kernels.} \emph{Insurance: Mathematics and Economics}, 105, 54--63.
\bibAnnoteFile{ang2022hierarchical}

\bibitem[{Bastos et~al.(2019)Bastos, Economou, Gomes, Villela, Coelho, Cruz, Stoner, Bailey, and Code{\c{c}}o}]{bstos2019reporting}
Bastos, Leonardo~S, Theodoros Economou, Marcelo~FC Gomes, Daniel~AM Villela, Fl{\'a}vio~C Coelho, Oswaldo~G Cruz, Owen Stoner, Tom Bailey, and Claudia~T Code{\c{c}}o (2019), \enquote{A modelling approach for correcting reporting delays in disease surveillance data.} \emph{Statistics in Medicine}, 38, 4363--4377.
\bibAnnoteFile{bstos2019reporting}

\bibitem[{Brinks(2014)}]{brinks2014illness}
Brinks, Ralph (2014), \enquote{Illness--death model: an efficient tool for non-parametric analysis of chronic disease epidemiology.} \emph{BMC Public Health}, 14, 644.
\bibAnnoteFile{brinks2014illness}

\bibitem[{B{\"u}hlmann(1967)}]{Buhlmann1967}
B{\"u}hlmann, Hans (1967), \enquote{Experience rating and credibility.} \emph{ASTIN Bulletin}, 4, 199--207.
\bibAnnoteFile{Buhlmann1967}

\bibitem[{Cai et~al.(2024)Cai, Abdallah, and Jeganathan}]{cai24}
Cai, Pengfei, Anas Abdallah, and Pratheepa Jeganathan (2024), \enquote{Recurrent neural networks for multivariate loss reserving and risk capital analysis.} \emph{arXiv preprint arXiv:2402.10421}.
\bibAnnoteFile{cai24}

\bibitem[{Carstensen(2007)}]{carstensen2007age}
Carstensen, Bendix (2007), \enquote{Age--period--cohort models for the {Lexis} diagram.} \emph{Statistics in Medicine}, 26, 3018--3045.
\bibAnnoteFile{carstensen2007age}

\bibitem[{Croux et~al.(2007)Croux, Filzmoser, and Oliveira}]{cfo07pcagrid}
Croux, Christophe, Peter Filzmoser, and Maria~Rosario Oliveira (2007), \enquote{Algorithms for projection--pursuit robust principal component analysis.} \emph{Chemometrics and Intelligent Laboratory Systems}, 87, 218--225.
\bibAnnoteFile{cfo07pcagrid}

\bibitem[{Cummins et~al.(1992)Cummins, Doherty, and Lo}]{CumminsDohertyLo1992}
Cummins, J.~David, Neil~A. Doherty, and James Lo (1992), \enquote{The economics of insurance regulation.} \emph{Journal of Risk and Insurance}, 59, 623--654.
\bibAnnoteFile{CumminsDohertyLo1992}

\bibitem[{Cummins and Weiss(2000)}]{CumminsWeiss2000}
Cummins, J.~David and Mary~A. Weiss (2000), \enquote{The global market for reinsurance: Consolidation, capacity, and efficiency.} \emph{Brookings-Wharton Papers on Financial Services}, 2000, 159--222.
\bibAnnoteFile{CumminsWeiss2000}

\bibitem[{Dai et~al.(2020)Dai, Mrkvi{\v{c}}ka, Sun, and Genton}]{dai2020}
Dai, Wenlin, Tom{\'a}{\v{s}} Mrkvi{\v{c}}ka, Ying Sun, and Marc~G Genton (2020), \enquote{Functional outlier detection and taxonomy by sequential transformations.} \emph{Computational Statistics \& Data Analysis}, 149, 106960.
\bibAnnoteFile{dai2020}

\bibitem[{Dawid(1984)}]{Dawid1984}
Dawid, A.~Philip (1984), \enquote{Present position and potential developments: Some personal views: Statistical theory: The prequential approach.} \emph{Journal of the Royal Statistical Society: Series A}, 147, 278--292.
\bibAnnoteFile{Dawid1984}

\bibitem[{De~Angelis and Gilks(1994)}]{Angelis2018}
De~Angelis, D. and W.~R. Gilks (1994), \enquote{Estimating acquired immune deficiency syndrome incidence accounting for reporting delay.} \emph{Journal of the Royal Statistical Society Series A: Statistics in Society}, 157, 31--40.
\bibAnnoteFile{Angelis2018}

\bibitem[{Denuit et~al.(2021)Denuit, Charpentier, and Trufin}]{denuit2021autocalibration}
Denuit, Michel, Arthur Charpentier, and Julien Trufin (2021), \enquote{Autocalibration and {T}weedie-dominance for insurance pricing with machine learning.} \emph{Insurance: Mathematics and Economics}, 101, 485--497.
\bibAnnoteFile{denuit2021autocalibration}

\bibitem[{Diebold and Mariano(1995)}]{DieboldMariano1995}
Diebold, Francis~X and Roberto~S Mariano (1995), \enquote{Comparing predictive accuracy.} \emph{Journal of Business \& Economic Statistics}, 13, 253--263.
\bibAnnoteFile{DieboldMariano1995}

\bibitem[{Elías et~al.(2022)Elías, Jiménez, and Shang}]{Elias2022}
Elías, Antonio, Raúl Jiménez, and Han~Lin Shang (2022), \enquote{On projection methods for functional time series forecasting.} \emph{Journal of Multivariate Analysis}, 189, 104890.
\bibAnnoteFile{Elias2022}

\bibitem[{Gabrielli(2021)}]{gabrielli21}
Gabrielli, Andrea (2021), \enquote{An individual claims reserving model for reported claims.} \emph{European Actuarial Journal}, 11, 541--577.
\bibAnnoteFile{gabrielli21}

\bibitem[{Gan and Lin(2015)}]{GanLin2015IME}
Gan, Guojun and X.~Sheldon Lin (2015), \enquote{Valuation of large variable-annuity portfolios under nested simulation: a functional data approach.} \emph{Insurance: Mathematics and Economics}, 62, 138--150.
\bibAnnoteFile{GanLin2015IME}

\bibitem[{Gneiting and Raftery(2007)}]{GneitingRaftery2007}
Gneiting, Tilmann and Adrian~E. Raftery (2007), \enquote{Strictly proper scoring rules, prediction, and estimation.} \emph{Journal of the American Statistical Association}, 102, 359--378.
\bibAnnoteFile{GneitingRaftery2007}

\bibitem[{Graziadei et~al.(2023)Graziadei, de~Melo, Targino et~al.}]{graziadei2023conformal}
Graziadei, Helton, Eduardo~FL de~Melo, Rodrigo~S Targino, et~al. (2023), \enquote{Conformal prediction for frequency-severity modeling.} \emph{arXiv preprint arXiv:2307.13124}.
\bibAnnoteFile{graziadei2023conformal}

\bibitem[{Griffith(2025)}]{griffith2025multioutput}
Griffith, Devan (2025), \enquote{Multioutput {G}aussian processes for loss ratio development.} \emph{Variance}, 18.
\bibAnnoteFile{griffith2025multioutput}

\bibitem[{Harrington and Niehaus(2000)}]{HarringtonNiehaus2000}
Harrington, Scott~E. and Greg Niehaus (2000), \enquote{Volatility and underwriting cycles.} \emph{Handbook of Insurance}, 657--686.
\bibAnnoteFile{HarringtonNiehaus2000}

\bibitem[{H{\"o}hle(2017)}]{hohle2017surveillance}
H{\"o}hle, Michael (2017), \enquote{An {R} package for the analysis of infectious disease surveillance data.} \emph{Journal of Statistical Software}, 77, 1--54.
\bibAnnoteFile{hohle2017surveillance}

\bibitem[{H{\"o}hle and an~der Heiden(2014)}]{hohle2014bayesian}
H{\"o}hle, Michael and Matthias an~der Heiden (2014), \enquote{Bayesian nowcasting during the {STEC} o104:{H4} outbreak in {G}ermany, 2011.} \emph{Biometrics}, 70, 993--1002.
\bibAnnoteFile{hohle2014bayesian}

\bibitem[{Hong(2025)}]{hong2025conformal}
Hong, Liang (2025), \enquote{Conformal prediction of future insurance claims in the regression problem.} \emph{arXiv preprint arXiv:2503.03659}.
\bibAnnoteFile{hong2025conformal}

\bibitem[{Hunt and Villegas(2015)}]{HuntVillegas2015IME}
Hunt, Andrew and Andres~M. Villegas (2015), \enquote{Robustness and convergence in the {L}ee--{C}arter model with cohort effects.} \emph{Insurance: Mathematics and Economics}, 64, 186--202.
\bibAnnoteFile{HuntVillegas2015IME}

\bibitem[{Kuo(2019)}]{kuo2019}
Kuo, Kevin (2019), \enquote{Deeptriangle: A deep learning approach to loss reserving.} \emph{Risks}, 7, 97.
\bibAnnoteFile{kuo2019}

\bibitem[{Lally and Hartman(2018)}]{lally2018estimating}
Lally, Nathan and Brian Hartman (2018), \enquote{Estimating loss reserves using hierarchical {B}ayesian {G}aussian process regression with input warping.} \emph{Insurance: Mathematics and Economics}, 82, 124--140.
\bibAnnoteFile{lally2018estimating}

\bibitem[{Lawless(1994)}]{lawless1994adjustments}
Lawless, Jerald~F (1994), \enquote{Adjustments for reporting delays and the prediction of occurred but not reported events.} \emph{Canadian Journal of Statistics}, 22, 15--31.
\bibAnnoteFile{lawless1994adjustments}

\bibitem[{Ludkovski and Zail(2022)}]{ludkovski2020gaussian}
Ludkovski, Mike and Howard Zail (2022), \enquote{Gaussian process models for incremental loss ratios.} \emph{Variance}, 15.
\bibAnnoteFile{ludkovski2020gaussian}

\bibitem[{Mack(1993)}]{m93}
Mack, Thomas (1993), \enquote{Distribution-free calculation of the standard error of chain ladder reserve estimates.} \emph{ASTIN Bulletin}, 23, 213--225.
\bibAnnoteFile{m93}

\bibitem[{Narisetty and Nair(2016)}]{extremaldepth}
Narisetty, Naveen~N. and Vijayan~N. Nair (2016), \enquote{Extremal depth for functional data and applications.} \emph{Journal of the American Statistical Association}, 111, 1705--1714.
\bibAnnoteFile{extremaldepth}

\bibitem[{Nieto-Barajas and Targino(2021)}]{nieto2021gamma}
Nieto-Barajas, Luis~E and Rodrigo~S Targino (2021), \enquote{A {G}amma moving average process for modelling dependence across development years in run-off triangles.} \emph{ASTIN Bulletin: The Journal of the IAA}, 51, 245--266.
\bibAnnoteFile{nieto2021gamma}

\bibitem[{Patton(2011)}]{Patton2011}
Patton, Andrew~J (2011), \enquote{Volatility forecast comparison using imperfect volatility proxies.} \emph{Journal of Econometrics}, 160, 246--256.
\bibAnnoteFile{Patton2011}

\bibitem[{Radtke et~al.(2016)Radtke, Schmidt, and Schnaus}]{rka16}
Radtke, Michael, Klaus~D. Schmidt, and Anja Schnaus (2016), \emph{Handbook on Loss Reserving}. Springer.
\bibAnnoteFile{rka16}

\bibitem[{Ramsay and Silverman(2005)}]{ramsay2005functional}
Ramsay, J.~O. and B.~W. Silverman (2005), \emph{Functional Data Analysis}, 2nd edition. Springer.
\bibAnnoteFile{ramsay2005functional}

\bibitem[{Rousseeuw and Croux(1993)}]{rc93}
Rousseeuw, Peter~J and Christophe Croux (1993), \enquote{Alternatives to the median absolute deviation.} \emph{Journal of the American Statistical Association}, 88, 1273--1283.
\bibAnnoteFile{rc93}

\bibitem[{Schneider and Schwab(2025)}]{schneider2025advancing}
Schneider, Judith~C and Brandon Schwab (2025), \enquote{Advancing loss reserving: A hybrid neural network approach for individual claim development prediction.} \emph{Journal of Risk and Insurance}, 92, 389--423.
\bibAnnoteFile{schneider2025advancing}

\bibitem[{Shang(2013)}]{shang13}
Shang, Han~Lin (2013), \enquote{Functional time series approach for forecasting very short-term electricity demand.} \emph{Journal of Applied Statistics}, 40.
\bibAnnoteFile{shang13}

\bibitem[{Shang(2017)}]{shang2017forecasting}
Shang, Han~Lin (2017), \enquote{Forecasting intraday {S\&P} 500 index returns: A functional time series approach.} \emph{Journal of forecasting}, 36, 741--755.
\bibAnnoteFile{shang2017forecasting}

\bibitem[{Shang and Hyndman(2011)}]{shang2011nonparametric}
Shang, Han~Lin and Rob~J Hyndman (2011), \enquote{Nonparametric time series forecasting with dynamic updating.} \emph{Mathematics and Computers in Simulation}, 81, 1310--1324.
\bibAnnoteFile{shang2011nonparametric}

\bibitem[{Shi(2017)}]{shi17}
Shi, Peng (2017), \enquote{A multivariate analysis of intercompany loss triangles.} \emph{Journal of Risk and Insurance}, 84, 717--737.
\bibAnnoteFile{shi17}

\bibitem[{Shi and Frees(2011)}]{sf11}
Shi, Peng and Edward~W Frees (2011), \enquote{Dependent loss reserving using copulas.} \emph{ASTIN Bulletin: The Journal of the IAA}, 41, 449--486.
\bibAnnoteFile{sf11}

\bibitem[{Shi and Hartman(2016)}]{sh16}
Shi, Peng and Brian~M Hartman (2016), \enquote{Credibility in loss reserving.} \emph{North American Actuarial Journal}, 20, 114--132.
\bibAnnoteFile{sh16}

\bibitem[{Shi and Shi(2023)}]{shi2023non}
Shi, Peng and Kun Shi (2023), \enquote{Non-life insurance risk classification using categorical embedding.} \emph{North American Actuarial Journal}, 27, 579--601.
\bibAnnoteFile{shi2023non}

\bibitem[{Steinmetz and Jentsch(2024)}]{steinmetz2024bootstrap}
Steinmetz, Julia and Carsten Jentsch (2024), \enquote{Bootstrap consistency for the {M}ack bootstrap.} \emph{Insurance: Mathematics and Economics}, 115, 83--121.
\bibAnnoteFile{steinmetz2024bootstrap}

\bibitem[{Stoner and Economou(2020)}]{stoner2020hierarchical}
Stoner, Owen and Theodoros Economou (2020), \enquote{A hierarchical framework for correcting reporting delays in disease surveillance data.} \emph{Biostatistics}, 21, 555--572.
\bibAnnoteFile{stoner2020hierarchical}

\bibitem[{Sun et~al.(2012)Sun, Genton, and Nychka}]{sun2012exact}
Sun, Ying, Marc~G Genton, and Douglas~W Nychka (2012), \enquote{Exact fast computation of band depth for large functional datasets: How quickly can one million curves be ranked?} \emph{Stat}, 1, 68--74.
\bibAnnoteFile{sun2012exact}

\bibitem[{Taylor(2019)}]{taylor19}
Taylor, Greg (2019), \enquote{Loss reserving models: Granular and machine learning forms.} \emph{Risks}, 7, 82.
\bibAnnoteFile{taylor19}

\bibitem[{Tibshirani(1996)}]{Tibshirani1996}
Tibshirani, Robert (1996), \enquote{Regression shrinkage and selection via the lasso.} \emph{Journal of the Royal Statistical Society: Series B}, 58, 267--288.
\bibAnnoteFile{Tibshirani1996}

\bibitem[{Tsai and Kim(2022)}]{TsaiKim2022IME}
Tsai, Chung-Cheng~Lee and Sojung Kim (2022), \enquote{Model mortality rates using property-and-casualty insurance reserving methods.} \emph{Insurance: Mathematics and Economics}, 106, 326--340.
\bibAnnoteFile{TsaiKim2022IME}

\bibitem[{Weiss(2001)}]{Weiss2001}
Weiss, Mary~A. (2001), \enquote{Factors affecting workers' compensation medical costs.} \emph{Journal of Risk and Insurance}, 68, 365--384.
\bibAnnoteFile{Weiss2001}

\bibitem[{Wilks(1990)}]{wilks1990combination}
Wilks, Daniel~S (1990), \enquote{On the combination of forecast probabilities for consecutive precipitation periods.} \emph{Weather and forecasting}, 5, 640--650.
\bibAnnoteFile{wilks1990combination}

\bibitem[{W{\"u}thrich(2018)}]{wuthrich18}
W{\"u}thrich, Mario~V (2018), \enquote{Machine learning in individual claims reserving.} \emph{Scandinavian Actuarial Journal}, 2018, 465--480.
\bibAnnoteFile{wuthrich18}

\bibitem[{W{\H{u}}thrich and Merz(2008)}]{wm08}
W{\H{u}}thrich, Mario~V. and Michael Merz (2008), \emph{Stochastic Claims Reserving Methods in Insurance}, volume 435. John Wiley and Sons.
\bibAnnoteFile{wm08}

\bibitem[{Zhang and Dukic(2013)}]{zd13}
Zhang, Yanwei and Vanja Dukic (2013), \enquote{Predicting multivariate insurance loss payments under the {B}ayesian copula framework.} \emph{Journal of Risk and Insurance}, 80, 891--919.
\bibAnnoteFile{zd13}

\end{thebibliography}
	\bibliographystyle{te}

    
\appendix
\renewcommand{\thesection}{\Alph{section}}
\section{Appendix}

\subsection{Mathematical Study of Functional Outliers}\label{app:outliers}

For comparison, we carry out outlier detection using the alternative concept of \emph{functional data depth}. This approach defines a metric that directly compares a given curve $y_{i,t}(x)$ to a collection of curves ${\cal F} = \{ {I}^m(x): m=1,\ldots,M\}$  in the sense of ``depth". The deepest curve is interpreted as the functional median, and the shallowest curves are outliers, being ``outside" or on the edge of the collection. 
In particular, we consider outliers and center curves defined in terms of Band Depth (BD), Modified Band Depth (MBD) and  Extremal Depth (EXD, \cite{extremaldepth}). 

Functional depth is a statistical framework for ranking functional observations from most outlying to most typical, providing a center-outward ordering on multivariate data.  While there is a canonical way to sort and rank real-valued scalars, there is a wide range of possible functional depths. The marginal rank $R(x)$ of $y_{i,t}$ at lag $x$ with respect to the collection $\mathcal{F}$ is
\begin{equation}
    R(x) := \sum_m 1_{\{I^{m}(x) < y_{i,t}(x) \} } \in \{1,\ldots,M\}.
\end{equation}

The (rank-2) Band Depth (BD) quantifies the frequency that $y_{i,t}$ falls into
the interval defined by another pair $y_{j,s}, y_{j', t'}$ in $\mathcal{F}$:
\begin{equation}\label{eq:bd}
    BD(y_{i,t}) := \frac{ \left( \min_x R(x) - 1 \right) \cdot \left( M - \max_x R(x) \right) + M - 1}{M (M + 1)}.
\end{equation}

Instead of looking only at the smallest and highest rank in Equation~\eqref{eq:bd}, Modified Band Width averages the product $(R(x)-1)(M-R(x))$ of the ranks of $y_{i,t}$ across all the 10 lags:
\begin{equation}\label{eq:MBD}
    MBD(y_{i,t}) := \frac{1}{M (M + 1)} \sum_{x=1}^{10} \frac{ \left( R(x) - 1 \right) \left( M - R(x) \right) }{M - 10 - 1}.
\end{equation}
The MBD is symmetric around the median and underlies the commonly used functional boxplot for identifying magnitude outliers. 

Extremal Depth (EXD), originally due to \cite{extremaldepth}, focuses on the extremal behavior of a curve, recording how \emph{often} it attains very high/low values relative to ${\cal F}$. To this end, EXD employs the point-wise depth 
\begin{equation}
    d(x) := 1 - \frac{\left| 2R(x) - M - 1\right|}{M}.
    \label{eq:EXD}
\end{equation}
Rather than integrating $d_\cdot$, EXD constructs the depth Cumulative Density Function (d-CDF)
\begin{equation}
    \Phi(r; y_{i,t}) := \frac{1}{10} \sum_{x=1}^{10} 1_{\{d(x) \leq r \}}, \qquad r \in (0,1)
\end{equation}
and then ranks an ILR curve $y_{i,t}$ according to the left-tail stochastic ordering of $\Phi$'s: 
\begin{equation}\label{eq:exd-defn}
    EXD(y_{i,t}) := \frac{1}{M} \sum_{m=1}^M 1_{\{ {\Phi} \succcurlyeq {\Phi}_{m} \}} ,
\end{equation}
where the stochastic order $\Phi\succcurlyeq \Phi_{m}$ intuitively means that the frequency (across lags $x$) of $y_{i,t}(\cdot)$ achieving an extreme rank is higher than that for $I^m(\cdot)$. 

For all of the above methods, the median curve in $\cal F$ is defined as the one with the highest depth, and the respective ranking also provides a natural notion of a functional quantile interval. 


All our analysis is performed within the \texttt{R} statistical environment. The RPCA analysis is done using \texttt{R} package \texttt{pcaPP} and functional depth using \texttt{fdaoutlier}. Visualization is done using \texttt{ggplot2} package. A GitHub repository containing the codes to replicate the result is available upon request.

\subsection{Additional Tables}
\label{app:tables}

\begin{table}[H]
\centering
\scriptsize
\setlength{\tabcolsep}{6pt}
\renewcommand{\arraystretch}{1.1}
\begin{tabular}{cc*{10}{S[table-format=-1.4]}}
\toprule
$s$ & $k$ & {PC0}   & {PC1}   & {PC2}   & {PC3}   & {PC4}   & {PC5}   & {PC6}   & {PC7}   & {PC8}   & {PC9}   \\
\midrule
1 & 1 & -0.5341 & -0.7187 & -0.3734 & -0.2053 & -0.0951 & -0.0677 & -0.0402 & -0.0259 & -0.0180 & -0.0189\\
2 & 1 & -0.5215 & -0.7230 & -0.3798 & -0.2093 & -0.0972 & -0.0691 & -0.0410 & -0.0264 & -0.0184 & -0.0192\\
3 & 1 & -0.5124 & -0.7225 & -0.3861 & -0.2177 & -0.1021 & -0.0724 & -0.0429 & -0.0278 & -0.0194 & -0.0199\\
3 & 2 & -0.7897 & 0.1979 & 0.3955 & 0.3352 & 0.2029 & 0.1328 & 0.0690 & 0.0509 & 0.0405 & 0.0248\\
3 & 3 & 0.3366 & -0.6424 & 0.3761 & 0.4290 & 0.2881 & 0.1895 & 0.1329 & 0.0856 & 0.0605 & 0.0305\\
\addlinespace
4 & 1 & -0.5034 & -0.7258 & -0.3911 & -0.2178 & -0.1031 & -0.0732 & -0.0431 & -0.0279 & -0.0195 & -0.0201\\
4 & 2 & -0.7968 & 0.1941 & 0.3911 & 0.3282 & 0.2005 & 0.1312 & 0.0678 & 0.0501 & 0.0400 & 0.0244\\
4 & 3 & 0.3332 & -0.6402 & 0.3847 & 0.4259 & 0.2892 & 0.1905 & 0.1333 & 0.0857 & 0.0606 & 0.0307\\
4 & 4 & 0.0177 & 0.1577 & -0.7325 & 0.4317 & 0.3929 & 0.2041 & 0.1686 & 0.1216 & 0.1015 & 0.0478\\
5 & 1 & -0.4813 & -0.7332 & -0.3988 & -0.2258 & -0.1065 & -0.0763 & -0.0446 & -0.0291 & -0.0205 & -0.0207\\
\addlinespace
5 & 2 & -0.8026 & 0.1612 & 0.3940 & 0.3220 & 0.2077 & 0.1339 & 0.0699 & 0.0514 & 0.0409 & 0.0251\\
5 & 3 & 0.3511 & -0.6454 & 0.4017 & 0.4021 & 0.2761 & 0.1834 & 0.1281 & 0.0818 & 0.0573 & 0.0294\\
5 & 4 & 0.0238 & 0.1396 & -0.7047 & 0.3696 & 0.4733 & 0.2297 & 0.1896 & 0.1346 & 0.1116 & 0.0579\\
5 & 5 & -0.0027 & -0.0098 & -0.1622 & 0.7164 & -0.6447 & -0.1293 & -0.1208 & -0.0702 & -0.0553 & -0.0743\\
6 & 1 & -0.4582 & -0.7381 & -0.4096 & -0.2333 & -0.1120 & -0.0801 & -0.0471 & -0.0309 & -0.0218 & -0.0217\\
\addlinespace
6 & 2 & -0.8254 & 0.1595 & 0.3683 & 0.3051 & 0.1982 & 0.1282 & 0.0663 & 0.0491 & 0.0394 & 0.0241\\
6 & 3 & 0.3283 & -0.6404 & 0.4139 & 0.4096 & 0.2828 & 0.1859 & 0.1308 & 0.0837 & 0.0587 & 0.0303\\
6 & 4 & 0.0276 & 0.1384 & -0.7087 & 0.3823 & 0.4591 & 0.2278 & 0.1885 & 0.1341 & 0.1112 & 0.0570\\
6 & 5 & 0.0024 & 0.0129 & 0.1472 & -0.7101 & 0.6531 & 0.1320 & 0.1250 & 0.0730 & 0.0576 & 0.0757\\
6 & 6 & 0.0076 & -0.0098 & -0.0055 & 0.1803 & 0.4644 & -0.7457 & -0.3049 & -0.2174 & -0.1752 & -0.1569\\
\addlinespace
7 & 1 & -0.4407 & -0.7430 & -0.4144 & -0.2386 & -0.1135 & -0.0863 & -0.0500 & -0.0331 & -0.0236 & -0.0230\\
7 & 2 & -0.8390 & 0.1537 & 0.3538 & 0.2930 & 0.1936 & 0.1245 & 0.0647 & 0.0478 & 0.0385 & 0.0235\\
7 & 3 & 0.3174 & -0.6371 & 0.4243 & 0.4074 & 0.2842 & 0.1918 & 0.1338 & 0.0857 & 0.0603 & 0.0318\\
7 & 4 & 0.0288 & 0.1335 & -0.7088 & 0.3897 & 0.4518 & 0.2242 & 0.1971 & 0.1347 & 0.1117 & 0.0570\\
7 & 5 & 0.0029 & 0.0179 & 0.1377 & -0.7129 & 0.6456 & 0.1455 & 0.1354 & 0.0782 & 0.0620 & 0.0793\\
\addlinespace
7 & 6 & 0.0118 & -0.0094 & 0.0001 & 0.1712 & 0.4823 & -0.7339 & -0.3128 & -0.2161 & -0.1745 & -0.1557\\
8 & 1 & -0.4172 & -0.7439 & -0.4256 & -0.2503 & -0.1207 & -0.0934 & -0.0544 & -0.0351 & -0.0257 & -0.0244\\
8 & 2 & -0.8526 & 0.1331 & 0.3383 & 0.2898 & 0.1853 & 0.1225 & 0.0608 & 0.0456 & 0.0371 & 0.0223\\
8 & 3 & 0.3129 & -0.6411 & 0.4210 & 0.4134 & 0.2858 & 0.1857 & 0.1265 & 0.0810 & 0.0575 & 0.0294\\
8 & 4 & 0.0284 & 0.1306 & -0.7194 & 0.4397 & 0.3912 & 0.2247 & 0.1904 & 0.1290 & 0.1095 & 0.0533\\
\addlinespace
8 & 5 & 0.0011 & 0.0250 & 0.0873 & -0.6822 & 0.6621 & 0.1954 & 0.1612 & 0.0970 & 0.0785 & 0.0905\\
8 & 6 & 0.0085 & 0.0024 & -0.0095 & 0.1469 & 0.5074 & -0.7499 & -0.2709 & -0.1935 & -0.1606 & -0.1479\\
8 & 7 & 0.0151 & -0.0014 & -0.0398 & 0.0225 & 0.1420 & 0.5134 & -0.7627 & -0.2758 & -0.2156 & -0.0981\\
8 & 8 & 0.0033 & -0.0034 & 0.0130 & -0.0074 & -0.0195 & -0.1314 & -0.4810 & 0.8093 & 0.2953 & 0.0931\\
9 & 1 & -0.3790 & -0.7542 & -0.4389 & -0.2515 & -0.1282 & -0.0959 & -0.0579 & -0.0366 & -0.0267 & -0.0257\\
\addlinespace
9 & 2 & -0.8715 & 0.1102 & 0.3064 & 0.2873 & 0.1737 & 0.1210 & 0.0599 & 0.0416 & 0.0382 & 0.0210\\
9 & 3 & 0.3061 & -0.6380 & 0.4560 & 0.4049 & 0.2772 & 0.1688 & 0.1169 & 0.0737 & 0.0490 & 0.0257\\
9 & 4 & 0.0535 & 0.1067 & -0.7059 & 0.4696 & 0.3902 & 0.2208 & 0.1867 & 0.1310 & 0.1061 & 0.0514\\
9 & 5 & -0.0082 & 0.0217 & 0.0728 & -0.6584 & 0.7049 & 0.1469 & 0.1536 & 0.0779 & 0.0766 & 0.0816\\
9 & 6 & 0.0027 & 0.0085 & -0.0167 & 0.1855 & 0.4653 & -0.7146 & -0.3575 & -0.2096 & -0.1991 & -0.1634\\
\addlinespace
9 & 7 & 0.0160 & -0.0007 & -0.0358 & 0.0015 & 0.1024 & 0.5825 & -0.7338 & -0.2648 & -0.1827 & -0.0824\\
9 & 8 & 0.0016 & -0.0039 & 0.0176 & -0.0144 & -0.0194 & -0.1233 & -0.4816 & 0.7963 & 0.3293 & 0.0966\\
\bottomrule
\end{tabular}
\caption{PCA Loadings by Component (PC0–PC9)}
\label{tab:pca_loadings}
\end{table}

\begin{table}[H]
\centering
\scriptsize
\setlength{\tabcolsep}{5pt}    
\renewcommand{\arraystretch}{1.1}  
\begin{tabular}{
  rr    
  *{2}{S[table-format=1.4]}   
  *{2}{S[table-format=1.4]}   
  *{4}{S[table-format=1.4]}   
  *{3}{S[table-format=1.4]}   
}
\toprule
\multirow{2}{*}{\textbf{s}}
  & \multirow{2}{*}{\textbf{k}}
  & \multicolumn{2}{c}{\textbf{Business Focus}}
  & \multicolumn{2}{c}{\textbf{NAIC Ownership}}
  & \multicolumn{4}{c}{\textbf{Geographic Focus}}
  & \multicolumn{3}{c}{\textbf{Time-variant Features}} \\
\cmidrule(lr){3-4} \cmidrule(lr){5-6} \cmidrule(lr){7-10} \cmidrule(lr){11-13}
 & & {personal} & {wkcomp} & {stock} & {other}
   & {midwest} & {northeast} & {south} & {west}
   & {time} & {log prem} & {log prem $\times$ time} \\
\midrule
1 & 1 & 0.0166 & -0.0026 & 0.0021 & -0.0304 & -0.0228 & -0.0062 & -0.0044 & 0.0112 & -0.0106 & -0.0085 & 0.0007\\
2 & 1 & 0.0159 & -0.0037 & 0.0014 & -0.0338 & -0.0206 & -0.0057 & -0.0005 & 0.0148 & -0.0110 & -0.0089 & 0.0008\\
3 & 1 & 0.0154 & -0.0042 & 0.0005 & -0.0361 & -0.0185 & -0.0047 & 0.0012 & 0.0171 & -0.0126 & -0.0098 & 0.0009\\
3 & 2 & -0.0028 & 0.0019 & 0.0050 & -0.0044 & -0.0197 & -0.0002 & -0.0081 &        &        & 0.0025 &       \\
3 & 3 & 0.0017 & -0.0041 & 0.0017 & -0.0028 & -0.0119 & 0.0134 & -0.0111 & -0.0011 & -0.0003 & 0.0009 &       \\
\addlinespace
4 & 1 & 0.0150 & -0.0047 & -0.0005 & -0.0377 & -0.0144 & -0.0049 & 0.0025 & 0.0194 & -0.0134 & -0.0100 & 0.0010\\
4 & 2 & -0.0029 & 0.0033 & 0.0049 & -0.0066 & -0.0199 & -0.0006 & -0.0095 & 0.0001 &        & 0.0026 &       \\
4 & 3 & 0.0013 & -0.0040 & 0.0023 & -0.0023 & -0.0121 & 0.0143 & -0.0107 & -0.0020 & -0.0004 & 0.0010 &       \\
4 & 4 &        & -0.0014 & -0.0009 & 0.0003 & -0.0026 & 0.0020 & -0.0039 &        &        & 0.0001 &       \\
\addlinespace
5 & 1 & 0.0138 & -0.0060 & -0.0028 & -0.0390 & -0.0105 & -0.0018 & 0.0056 & 0.0216 & -0.0130 & -0.0091 & 0.0009\\
5 & 2 & -0.0037 & 0.0032 & 0.0055 & -0.0088 & -0.0198 & 0.0012 & -0.0095 & 0.0027 &        & 0.0028 &       \\
5 & 3 & 0.0016 & -0.0032 & 0.0027 & -0.0020 & -0.0123 & 0.0137 & -0.0105 & -0.0030 & -0.0004 & 0.0009 &       \\
5 & 4 & -0.0008 & -0.0024 & -0.0017 & 0.0013 & -0.0045 & 0.0030 & -0.0049 & -0.0005 & -0.0014 & -0.0004 & 0.0001\\
5 & 5 &        & 0.0004 &        &        &        &        &        &        &        & -0.0004 &       \\
\addlinespace
6 & 1 & 0.0134 & -0.0084 & -0.0054 & -0.0416 & -0.0074 & 0.0023 & 0.0080 & 0.0245 & -0.0114 & -0.0077 & 0.0008\\
6 & 2 & -0.0042 & 0.0030 & 0.0051 & -0.0128 & -0.0190 & 0.0007 & -0.0092 & 0.0061 &        & 0.0027 & .0000\\
6 & 3 & 0.0010 & -0.0029 & 0.0031 & -0.0034 & -0.0138 & 0.0139 & -0.0114 & -0.0030 & -0.0004 & 0.0007 & .0000\\
6 & 4 & -0.0012 & -0.0028 & -0.0020 & 0.0013 & -0.0052 & 0.0035 & -0.0045 & -0.0008 & -0.0020 & -0.0009 & 0.0001\\
6 & 5 &        & -0.0005 &        &        &        &        &        &        &        & 0.0005 &       \\
6 & 6 & 0.0006 & 0.0006 & 0.0021 & 0.0011 & 0.0026 & 0.0025 & 0.0029 & -0.0005 &        & -0.0009 &       \\
\addlinespace
7 & 1 & 0.0132 & -0.0077 & -0.0078 & -0.0428 & -0.0028 & 0.0056 & 0.0086 & 0.0226 &        & -0.0008 & 0.0001\\
7 & 2 & -0.0043 & 0.0032 & 0.0040 & -0.0169 & -0.0184 & 0.0006 & -0.0099 & 0.0099 & 0.0036 & 0.0048 & -0.0002\\
7 & 3 & 0.0011 & -0.0017 & 0.0036 & -0.0040 & -0.0148 & 0.0131 & -0.0124 & -0.0033 & -0.0005 & 0.0009 &       \\
7 & 4 &        & -0.0021 & -0.0006 &        & -0.0044 & 0.0035 & -0.0027 &        &        & 0.0001 &       \\
7 & 5 &        & -0.0004 &        &        &        &        &        &        &        & 0.0005 &       \\
7 & 6 &        &        &        &        &        &        & 0.0004 &        &        & -0.0008 &       \\
\addlinespace
8 & 1 & 0.0142 & -0.0047 & -0.0094 & -0.0411 &        & 0.0066 & 0.0089 & 0.0189 &        & -0.0009 &       \\
8 & 2 & -0.0037 & 0.0035 & 0.0028 & -0.0203 & -0.0193 & 0.0014 & -0.0100 & 0.0115 & 0.0053 & 0.0055 & -0.0004\\
8 & 3 &        & -0.0012 & 0.0031 & -0.0038 & -0.0140 & 0.0123 & -0.0113 & -0.0011 & -0.0005 & 0.0005 &       \\
8 & 4 &        & -0.0014 &        &        & -0.0038 & 0.0033 & -0.0020 &        &        &        &       \\
8 & 5 &        & -0.0012 &        &        &        &        &        &        &        & 0.0006 &       \\
8 & 6 &        &        &        &        & 0.0003 &        & 0.0012 &        &        & -0.0006 &       \\
8 & 7 &        &        &        &        &        & 0.0014 &        &        &        & -0.0008 &       \\
8 & 8 &        &        & 0.0007 &        & -0.0003 & -0.0021 & 0.0002 &        &        & 0.0002 &       \\
\addlinespace
9 & 1 & 0.0147 &        & -0.0105 & -0.0335 &        & 0.0077 & 0.0053 & 0.0129 & 0.0006 & -0.0015 &       \\
9 & 2 & -0.0023 & 0.0050 & 0.0013 & -0.0225 & -0.0191 & 0.0028 & -0.0108 & 0.0108 & 0.0088 & 0.0070 & -0.0006\\
9 & 3 &        & -0.0008 & 0.0035 & -0.0053 & -0.0146 & 0.0121 & -0.0118 & -0.0011 & -0.0005 & 0.0005 &       \\
9 & 4 &        & -0.0008 &        &        & -0.0044 & 0.0039 & -0.0018 &        &        &        &       \\
9 & 5 &        &        &        &        &        &        &        &        &        & 0.0004 &       \\
9 & 6 &        & 0.0002 & 0.0010 & 0.0003 & 0.0024 & 0.0019 & 0.0028 &        &        & -0.0008 &       \\
9 & 7 &        &        &        &        &        & 0.0011 &        &        &        & -0.0005 &       \\
9 & 8 &        &        & 0.0009 &        &        & -0.0019 &        &        &        & 0.0002 &       \\
\bottomrule
\end{tabular}
\caption{LASSO Coefficients by Variable. Each row represents the coefficient estimates for a given development lag $s$ and factor index $k$.}
\label{tab:lasso_full}
\end{table}

\begin{table}[H]
    \centering
    \begin{tabular}{c|cc|ccc|ccc}
        \toprule
        Lag 
        & \multicolumn{2}{c|}{MAPE} 
        & \multicolumn{3}{c|}{Coverage \%} 
        & \multicolumn{3}{c}{Interval score} \\
        $s$ & PLS & CL & PLS & EXD & CL & PLS & EXD & CL \\
        \midrule
        1 & 21.84\% & 16.53\% & 95.62\% & 90.51\% & 86.13\% & 0.6684 & 0.9469 & 0.7098 \\
        2 & 13.18\% & 10.31\% & 97.81\% & 97.08\% & 79.56\% & 0.5130 & 0.4494 & 1.0422 \\
        3 & 9.69\%  & 7.21\%  & 99.27\% & 97.08\% & 71.53\% & 0.3675 & 0.3097 & 0.9904 \\
        4 & 6.12\%  & 5.18\%  & 100.00\% & 99.27\% & 70.80\% & 0.2590 & 0.2142 & 0.4047 \\
        5 & 4.08\%  & 4.10\%  & 100.00\% & 97.08\% & 73.72\% & 0.2037 & 0.1783 & 0.1808 \\
        6 & 2.52\%  & 2.69\%  & 100.00\% & 98.54\% & 79.56\% & 0.1529 & 0.1253 & 0.1176 \\
        7 & 2.09\%  & 2.07\%  & 100.00\% & 100.00\% & 84.67\% & 0.1090 & 0.1096 & 0.0900 \\
        8 & 1.33\%  & 1.41\%  & 100.00\% & 99.27\% & 83.94\% & 0.0766 & 0.0709 & 0.0669 \\
        9 & 0.70\%  & 0.67\%  & 95.62\% & 95.62\% & 92.70\% & 0.0334 & 0.0334 & 0.0354 \\
        \bottomrule
    \end{tabular}
    \caption{Statistical scores for forecasting ultimate cumulative losses without outliers removed. We report MAPE, coverage at 95\% level, and interval scores $IS_{\alpha}$ with $\alpha=0.95$. PLS refers to point-wise interval, CL to chain ladder and EXD to extremal depth.}
    \label{tbl:mape-coverage-withoutlier}
\end{table}

\begin{table}[H]
    \centering
    \begin{tabular}{c|ccc|ccc}
        \toprule
        \multirow{2}{*}{$s$} 
        & \multicolumn{3}{c|}{Coverage \%} 
        & \multicolumn{3}{c}{Interval Score} \\
        & PLS & EXD & CL & PLS & EXD & CL \\
        \midrule
        1 & 69.34\% & 75.18\% & 82.48\% & 0.4728 & 0.5759 & 0.5125 \\
        2 & 94.16\% & 95.62\% & 72.26\% & 0.3651 & 0.3429 & 0.4802 \\
        3 & 94.16\% & 94.16\% & 64.23\% & 0.2538 & 0.2346 & 0.4728 \\
        4 & 97.08\% & 97.81\% & 66.42\% & 0.1724 & 0.1575 & 0.2003 \\
        5 & 96.35\% & 97.08\% & 70.07\% & 0.1356 & 0.1349 & 0.1121 \\
        6 & 96.35\% & 97.08\% & 75.91\% & 0.1082 & 0.1011 & 0.0734 \\
        7 & 100.00\% & 100.00\% & 78.10\% & 0.0785 & 0.0854 & 0.0573 \\
        8 & 98.54\% & 99.27\% & 82.48\% & 0.0604 & 0.0617 & 0.0501 \\
        9 & 95.62\% & 95.62\% & 92.70\% & 0.0334 & 0.0334 & 0.0354 \\
        \bottomrule
    \end{tabular}
    \caption{Statistical scores for forecasting cumulative loss curves from given lag $s$ to lag 9 without outliers removed. We report coverage at 95\% level, and interval scores $IS_{\alpha}$ with $\alpha=0.95$. PLS refers to point-wise interval, CL to chain ladder and EXD to extremal depth.}
    \label{tbl:functional-score-withoutlier}
\end{table}

\subsection{Additional Plots}
\label{app:plots}

	\begin{figure}[!ht]
		\centering
		\begin{subfigure}[b]{0.45\textwidth}
			\includegraphics[width=\textwidth]{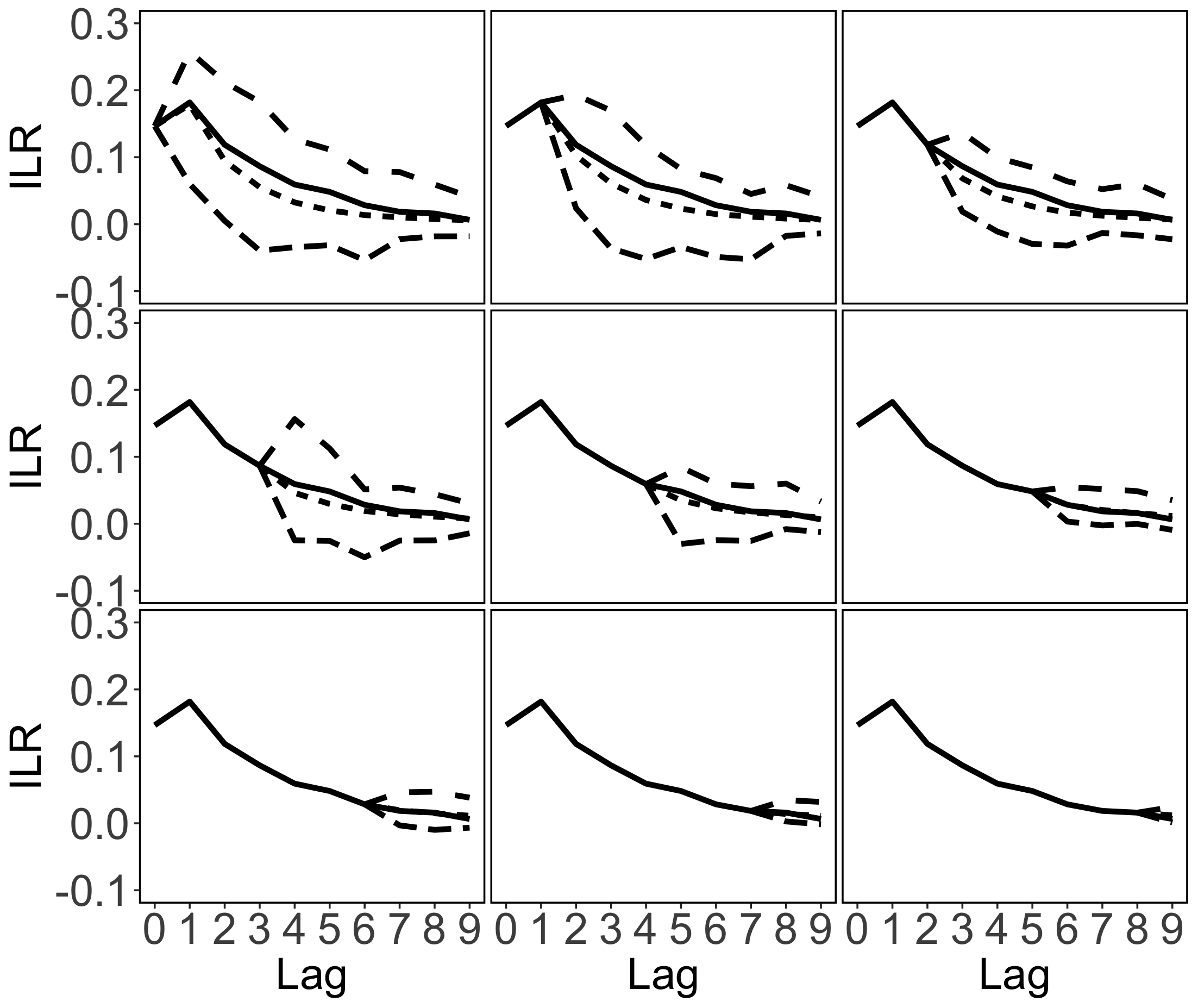}
			\caption{Company P53 ILR}
		\end{subfigure}
		\begin{subfigure}[b]{0.45\textwidth}
			\includegraphics[width=\textwidth]{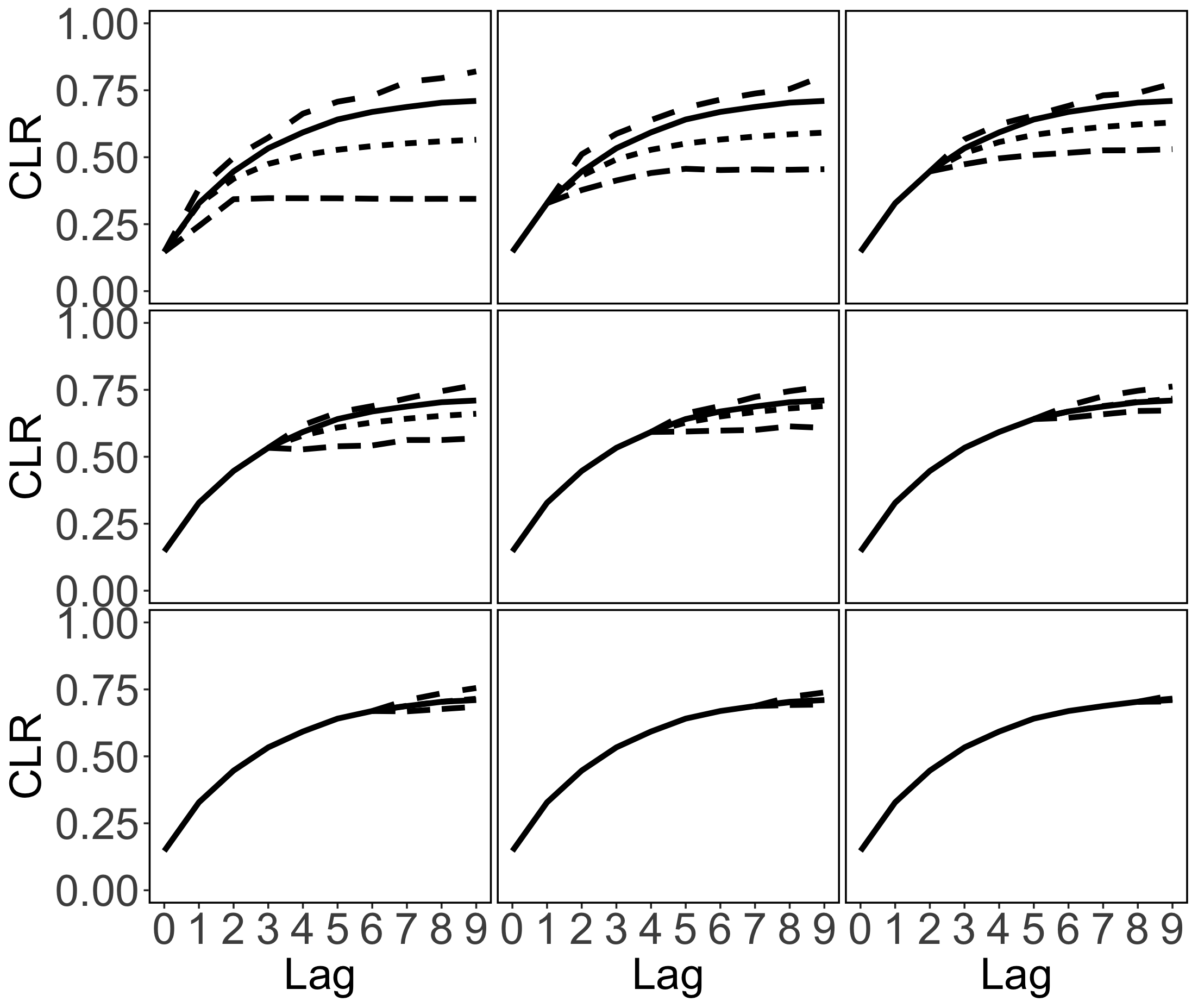}
			\caption{Company P53 CLR}
		\end{subfigure}
		
		\begin{subfigure}[b]{0.45\textwidth}
			\includegraphics[width=\textwidth]{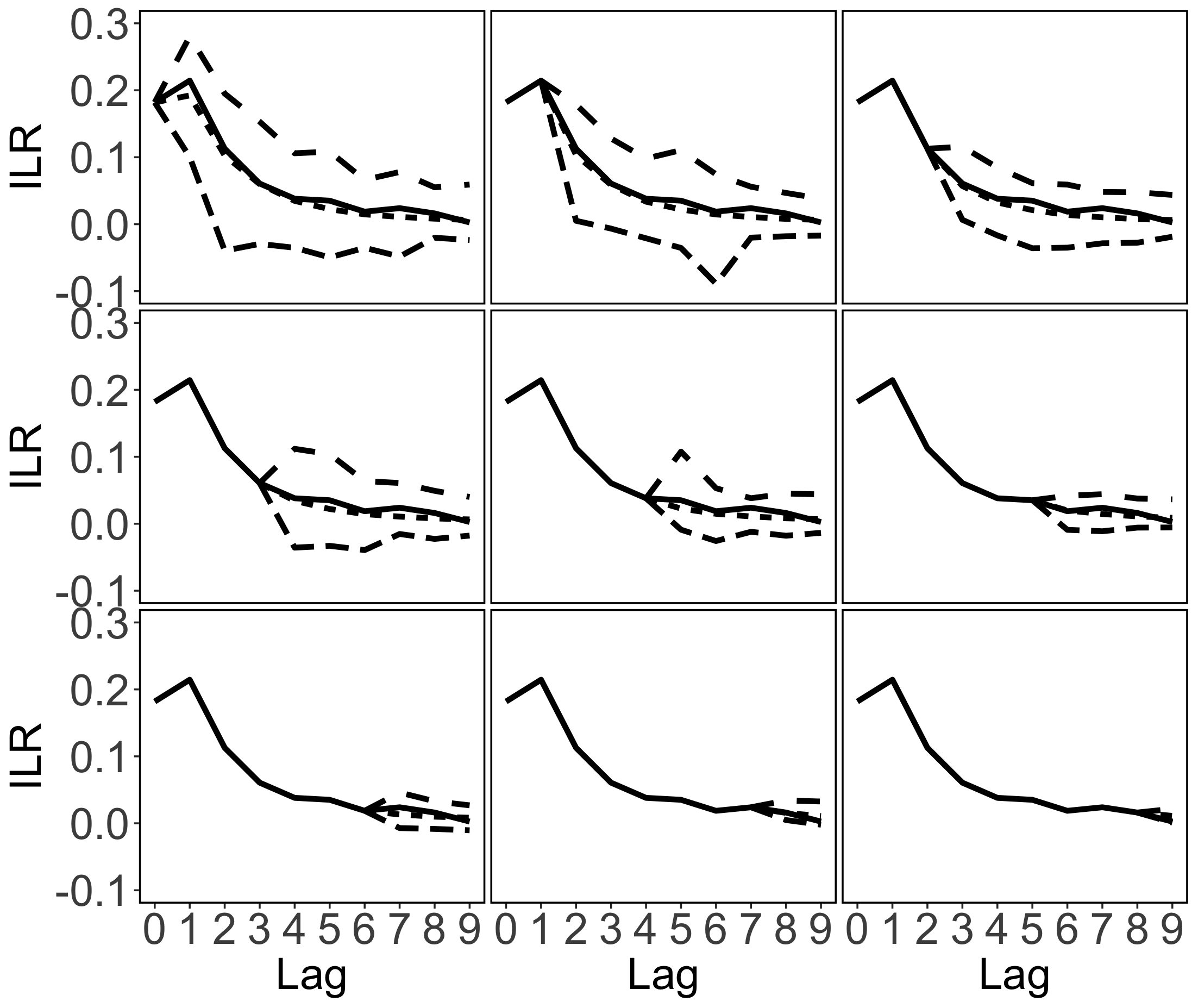}
			\caption{Company C260 ILR}
		\end{subfigure}
		\begin{subfigure}[b]{0.45\textwidth}
			\includegraphics[width=\textwidth]{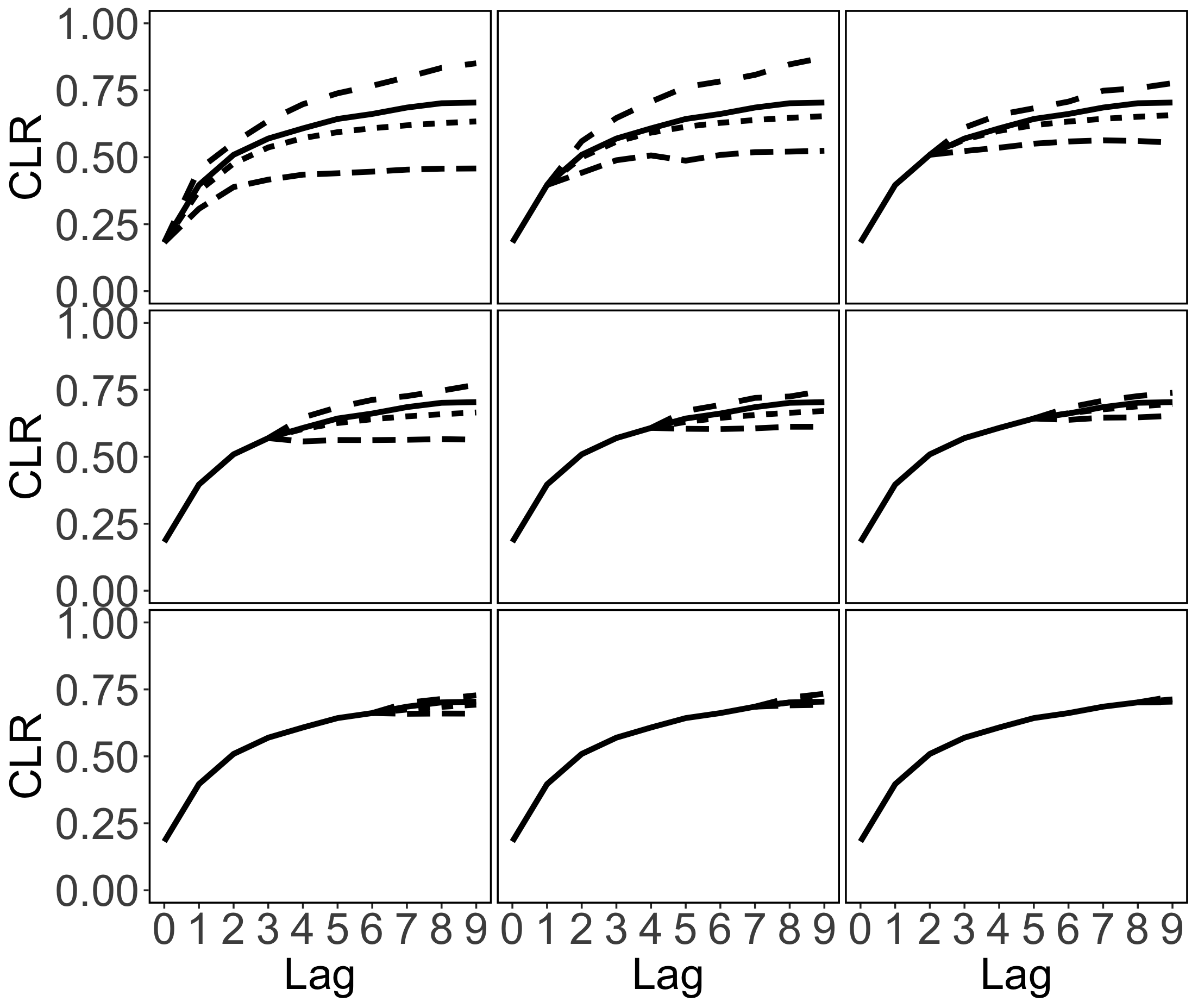}
			\caption{Company C260 CLR}
		\end{subfigure}

        \begin{subfigure}[b]{0.45\textwidth}
			\includegraphics[width=\textwidth]{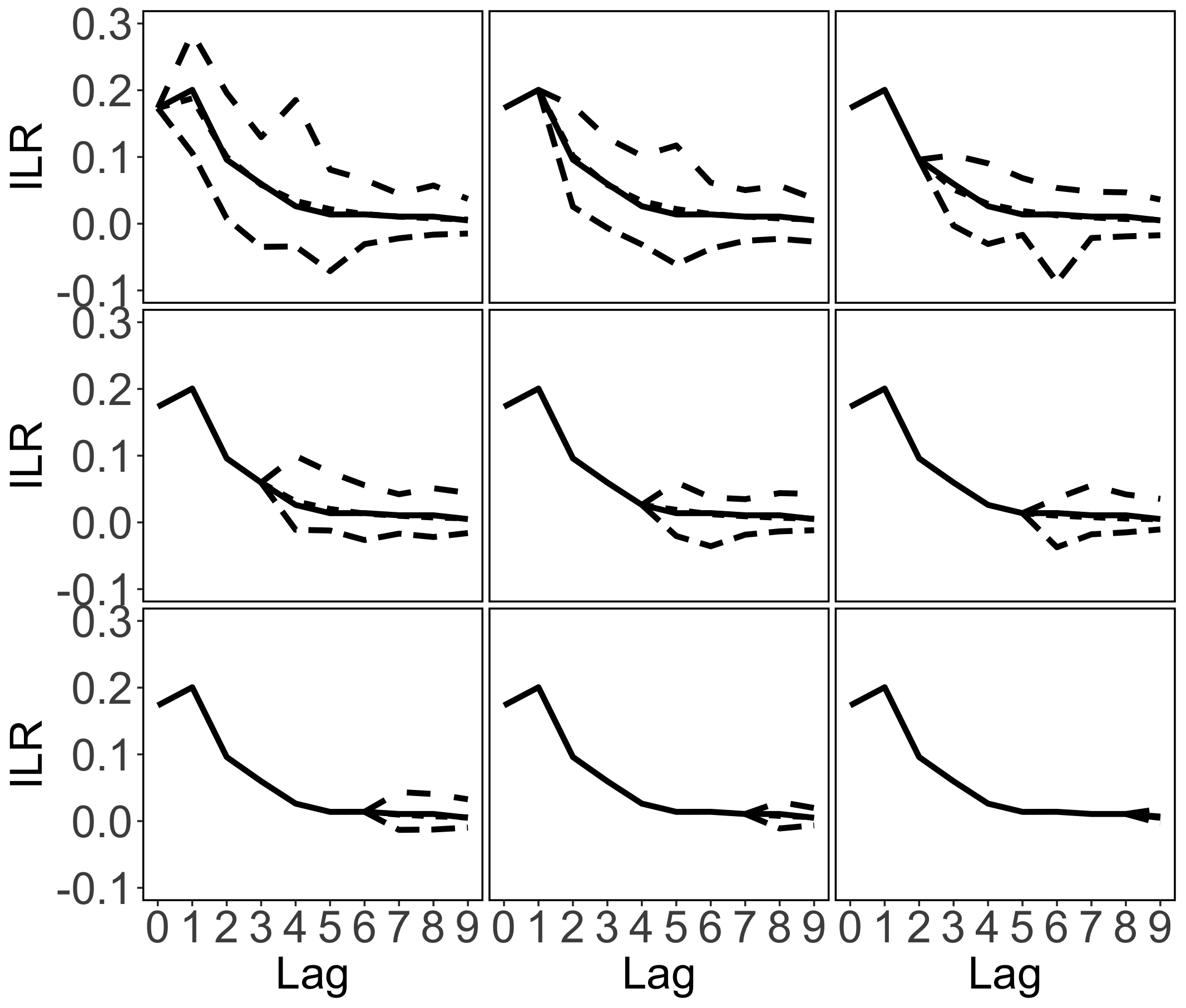}
			\caption{Company P1406 ILR}
		\end{subfigure}
		\begin{subfigure}[b]{0.45\textwidth}
			\includegraphics[width=\textwidth]{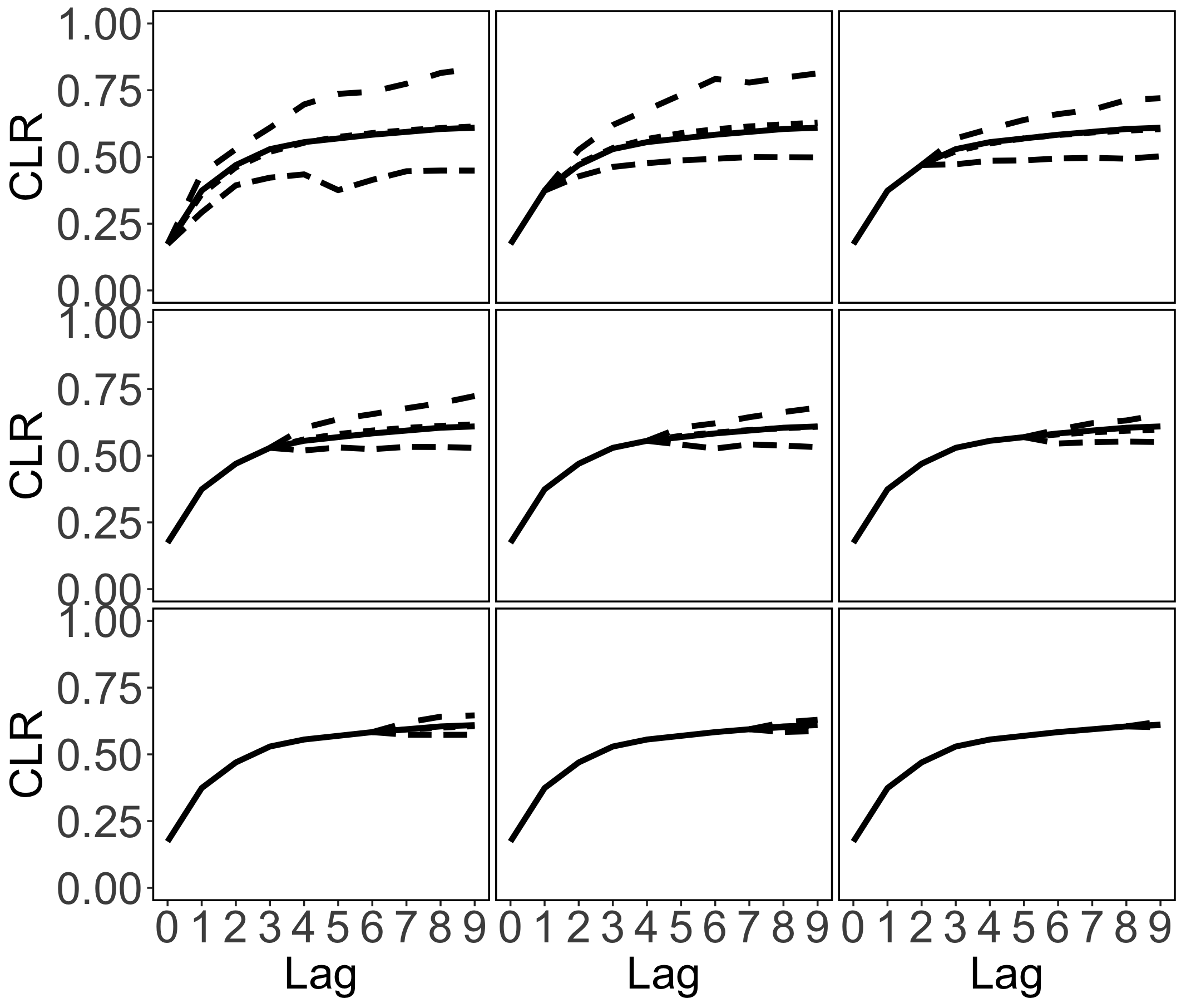}
			\caption{Company P1406 CLR}
		\end{subfigure}
		\caption{Forecast of ILR and CLR for AY 2010: solid line---actual ILR/CLR; short-dashed line---forecast ILR/CLR; long-dashed line---95\% bootstrap prediction interval }
        \label{fig:ilrclr2010} 
	\end{figure}

    \end{document}